\documentclass[12pt]{iopart}

\usepackage{iopams}
\usepackage{graphicx}

\newcommand{\bI}{\boldsymbol{I}}
\newcommand{\be}{\boldsymbol{e}}
\newcommand{\bq}{\boldsymbol{q}}
\newcommand{\bR}{\boldsymbol{R}}
\newcommand{\la}{\lambda}
\newcommand{\ka}{\kappa}
\newcommand{\kam}{\kappa_m}
\newcommand{\kat}{\kappa_T}
\newcommand{\kap}{\kappa_p}
\newcommand{\bD}{\boldsymbol{D}}
\newcommand{\mcn}{\mathcal{N}}
\newcommand{\mbfT}{\mathbf{T}}
\newcommand{\bk}{\boldsymbol{k}}
\newcommand{\mcs}{\mathcal{S}}
\newcommand{\mbfF}{\mathbf{F}}

\begin{document}

\title[Recasting Navier-Stokes]{Recasting Navier-Stokes Equations}

\author{M.~H.~Lakshminarayana Reddy$^1$\footnote{Author to whom any correspondence should be addressed.}, S.~Kokou~Dadzie$^1$,
  Raffaella Ocone$^1$,  Matthew~K.~Borg$^2$ and  Jason~M.~Reese$^2$}

\address{$^1$School of Engineering and Physical Sciences, Heriot-Watt University, Edinburgh EH14 4AS, Scotland, UK}
\address{$^2$School of Engineering, University of Edinburgh, Edinburgh EH9 3FB, UK}
\eads{\mailto{l.mh@hw.ac.uk} and \mailto{k.dadzie@hw.ac.uk}}
\vspace{10pt}

\begin{abstract}
Classical Navier-Stokes equations fail to describe some flows in both the compressible and incompressible configurations.
In this article, we propose a new methodology based on transforming the fluid mass velocity vector field to obtain a new class of continuum models. We uncover a class of continuum models which we call the re-casted Navier-Stokes. They naturally exhibit the physics of previously proposed models by different authors to substitute the original Navier-Stokes equations. The new models unlike the conventional Navier-Stokes appear as more complete forms of mass diffusion type continuum flow equations. They also form systematically a class of thermo-mechanically consistent hydrodynamic equations via the original equations. The plane wave analysis is performed to check their linear stability under small perturbations, which confirms that all re-casted models are spatially and temporally stable like their classical counterpart. We then use the Rayleigh-Brillouin scattering experiments to demonstrate that the re-casted equations may be better suited for explaining some of the experimental data where original Navier-Stokes fail.
\end{abstract}
\vspace{2pc}

%

\maketitle

%
%

\section{Introduction}
\label{1}

Fluid mechanics is one of the oldest field of science and still widely researched due to its broad spread of applications in many industries \cite{Durst2008}. The development of basic fundamental dynamic laws such as Newton's law of motion and Newton's law of viscosity culminated in the current form of the Navier-Stokes equations; these equations are still widely accepted as the universal basis of modelling fluid motion \cite{Darrigol2002,Darrigol2005}. They are frequently solved by using numerical computational methods. The classical Navier-Stokes equations are known to be inadequate in describing some compressible flows accurately \cite{Durst2008,Alsmeyer1976,Sone2000,Greenshields2007}. The failure may be tied up to the basic assumptions made while deriving them \cite{Stokes1845}. Improving the range of applicability of the Navier-Stokes equations beyond their limits has been and still is a critical area of research.

In the last decade, the classical Navier-Stokes equations have received a number of modifications/extensions based on the diffusion transport of mass \cite{Brenner200511,Brenner200560,Brenner2012,Ottinger2005,Durst2006,Durst2008,Sambasivam2012,Carrassi1972}. In particular, Brenner \cite{Brenner200511,Brenner200560} first proposed a set of modifications to the equations based on the physical arguments of thermophoretic motion in gases. In the end, he made a revision to the Newton's law of viscosity and the Fourier's law of heat conduction for compressible fluids. He argued that the velocities presented in the mass and momentum balance equations differ by a mass-diffusion flux term which he later adopted in his bi-velocity hydrodynamic theory \cite{Brenner2012}. Brenner's bi-velocity theory has received a fair amount of attention from the fluid mechanics research community due to its controversial nature that the classical Navier-Stokes equations could be incomplete or incorrect. Extended Navier-Stokes equations of Brenner type are criticized and rebutted in {\"O}ttinger \etal \cite{Ottinger2009} for not satisfying some combined mechanical properties. Following this, Dadzie \cite{Dadzie2013} derived a new set of continuum equations based on a Boltzmann-like kinetic equation that satisfies all thermo-mechanical properties. However, unlike the three classical conservation laws, Dadzie's model contains four set of transport equations, where the additional is a non-conservative equation for the volume transport.

{\"O}ttinger \cite{Ottinger2005} also proposed earlier a substitute for the classical Navier-Stokes in his phenomenological GENERIC formalism. In the GENERIC formulation, {\"O}ttinger \cite{Ottinger2005} demonstrated that by assuming the row and the column, which are associated with the mass density in the friction matrix to be identically zero, leads to the conventional Navier-Stokes equations. A more general form of friction matrix that includes the basic fact that particles operate diffusive motions, leads to a revised set of transport equations that contain two velocities, which are very similar in nature to the volume and mass velocities. Durst \etal. \cite{Durst2006} based their arguments on that the absence of mass diffusion terms in the continuity equation contradicts constitutive relations for momentum and heat diffusion: \emph{when the fluid properties changes spatially in the presence of momentum and heat diffusions, there should also be present the mass diffusion}. They later derived the Extended Navier-Stokes equations \cite{Durst2006,Sambasivam2012,Sambasivam2014} based on the mass-diffusion controlled formalism. A late suggestion to substitute the Navier-Stokes is given by Sv{\"a}rd \cite{Svard2018}.

Generally, there are a number of experimental data that standard Navier-Stokes fail to predict. Shock wave structure predictions are a ubiquitous example of Navier-Stokes failure \cite{Alsmeyer1976,Greenshields2007,Reddy2015,Reddy2016}. Experimental data of water flows in carbon nanotubes or confined pores is another research topic showing large deviations from the classical Hagen-Poiseuille equation \cite{Majumder2005,Ozerinc2010}. A convincing theoretical interpretation of this data is still lacking. Leaving aside non-linear configurations, conventional Navier-Stokes equations also fail to describe some of the linear flow problems accurately. For example, it is unsuccessful in describing the actual spectrum shapes in the Rayleigh-Brillouin scattering problem \cite{Mountain1966,Marques1993,Ma2012,Wu2014,Wu2015,Wu2018}.

In the current work, we provide new insights into the question of finding alternatives to the Navier-Stokes equations.
Starting with the basic conventional Navier-Stokes equations, we introduce a re-casting methodology. It involves transforming the fluid velocity field variable within the basic standard fluid flow equations with an appropriately selected \textit{change of variable}. Two linear problems are then considered to test our new derivations: sound wave dispersion in a monatomic gas and the Rayleigh-Brillouin light scattering experiments.
A better prediction of the experiments by the transformed equations (i.e., the models where the velocity vector field has an implicit diffusive component embedded) are observed. While the better prediction may not be trivial on the sound dispersion it is evident on the spontaneous Rayleigh-Brillouin scattering experiments. In a follow-up paper we show that the transformed equations perform better also on the description of shock wave profiles in monatomic gases (a nonlinear problem). These theoretical observations corroborate earlier observations of the difference between a dye- or photochromic experiments (measuring a fluid's mass velocity) and a tracer velocity \cite{HBrenner2009}.

The paper is organized as follows: in \S\ref{2} we derive three different new continuum models using, initially, three different change of variables. In the following section \S \ref{3}, linear stability analysis is performed for the new models. Then all re-casted models are applied to study the Rayleigh-Brillouin scattering problem and compared with the classical predictions and the experiments in \S \ref{4}. Finally, conclusions are drawn at the end.

\section{Theory}
\label{2}

Our new theory starts with the standard Navier-Stokes equations. That is the three standard conservation equations, closed using Newton's and Fourier's Laws representing the shear stress and the heat flux.

\subsection{Classical Navier-Stokes Equations}
\label{2.1}
The standard hydrodynamic equations are a differential form of three classical conservation laws.
Namely, mass, momentum and energy conservation laws that govern the motion of a fluid.
The classical Navier-Stokes equations in an Eulerian reference frame are given as \cite{Dadzie2013,Reddy2015}:

mass balance/continuity equation
\begin{equation}
\label{eqn_mass}
\frac{\partial \rho}{\partial t}  + \, \nabla \cdot [\rho\, U]  = 0,
\end{equation}

momentum balance equation
\begin{equation}
\label{eqn_momentum}
\frac{\partial  \rho \, U }{\partial t}  +  \, \nabla \cdot \left[\rho \,U\otimes\,U \right]\,+ \,\nabla \cdot \left[p \,\bI \,+ \,\bPi^{(NS)} \right] = 0,
\end{equation}

energy balance equation
\begin{eqnarray}
\label{eqn_energy}
&& \frac{\partial}{\partial t} \left[\frac{1}{2} \rho \,U^2\, +\, \rho \,\be_{in}\right] \, +\,\nabla \cdot \left[\frac{1}{2} \rho\, U^2\, U \,+\, \rho\, \be_{in}\, U\right]\,\nonumber \\
&& \qquad \qquad \qquad \qquad \,+\,\nabla \cdot \left[(p \,\pmb{I}\, + \,\bPi^{(NS)}) \cdot U \right] \,+\, \nabla \cdot \boldsymbol{q}^{(NS)} = 0,
\end{eqnarray}
where $\rho$ is the mass-density of the fluid, $U$ is the flow mass velocity, $p$ is the hydrostatic pressure and $\be_{in}$ is the specific internal energy of the fluid. Further, $\bPi^{(NS)}$ and $\bq^{(NS)}$ represent the shear stress tensor and the heat flux vector, respectively, and $\boldsymbol{I}$ is the identity matrix. All these hydrodynamic fields are functions of time $t$ and spatial variable $\boldsymbol{X}$. Here $\nabla$ and $\nabla \cdot$ denotes the usual spatial gradient operator and the divergence operator, respectively, and $U \otimes U$ represents the tensor product of two velocity vectors defined as in \eref{eqn_A2}. The expression for the  specific internal energy $\be_{in}$ is given by
\begin{equation}
 \be_{in} = \frac{p}{\rho (\gamma -1)},
\end{equation}
where $\gamma$ is the specific heat capacity ratio. The constitutive models for the shear stress $\bPi^{(NS)}$ and the heat flux vector $\boldsymbol{q}^{(NS)}$ are:
\begin{eqnarray}
\bPi^{(NS)} &= -2 \,\mu \left[ \frac{1}{2} (\nabla U \,+ \,\widetilde{\nabla U})\, -\, \frac{1}{3}\, \bI \,\left(\nabla \cdot U \right) \right] = -2\, \mu\, \mathring{\overline{\nabla U}}, \label{eqn_stress1}\\
\boldsymbol{q}^{(NS)} &= - \ka \, \nabla T, \label{eqn_heatflux1}
\end{eqnarray}
where $\nabla U$ is the spatial gradient of $U$ and $\widetilde{\nabla U}$ is the transpose of $\nabla U$. Coefficients $\mu$ and $\kappa$ are the dynamic viscosity and  the heat conductivity, respectively. The shear stress can be expressed in terms of the symmetric part of the velocity gradient and the divergence of the velocity field as
\begin{eqnarray}
\bPi^{(NS)} &= -2 \mu \left[ \mathbf{D}\left(U\right)\,-\,\frac{1}{3} \left( \nabla \cdot U\right)\, \bI \right]
= -\,2 \,\mu \,\mathbf{D}\left(U\right)\,-\,\la\, \left( \nabla \cdot U\right)\, \bI,
\end{eqnarray}
where $\la = -\frac{2}{3}\mu$ is the bulk-viscosity co-efficient.

The system \eref{eqn_mass} - \eref{eqn_heatflux1} is the well-known conventional fluid flow model for a viscous and heat conducting fluid. In the limit of vanishing viscous and heat conducting terms, these equations reduced to the well-known Euler equations, which are used to model inviscid and non-diffusive flows. It is trivial to observe in this system that continuity equation \eref{eqn_mass} does not contain a diffusion term, whereas the momentum and energy equations do. Hence, the classical Navier-Stokes equations form an incomplete parabolic system \cite{Belov1971}. In other words, they are prohibited from being fully parabolic due to the absence of diffusion term in the mass balance equation. In the meantime, this system can be shown to satisfy all required mechanical properties and also associates with a second law / entropy equation \cite{Ottinger2009,Woods1993}. It is also important to note that in constitutive equation \eref{eqn_stress1}, no complex contributions from effects such as fluid dilation, temperature gradient, fluid vorticity, etc. to the shear stress are described. It is this basic form of the classical fluid flow equations that are traditionally shown to satisfy all known thermo-mechanical properties \cite{Woods1993}.


\subsection{Re-casted Navier-Stokes equations - I (RNS - I): $U \rightarrow U_v-\kam \nabla \ln\rho$}
\label{2.2}

In this subsection we derive the first new set of hydrodynamic equations, by transforming the velocity field $U$ into a newly defined velocity field $U_v$. We assume that the flow mean mass velocity field $U$ can be written in terms of the new velocity field called the mean volume velocity $U_v$ \cite{Brenner200511,Brenner200560,DRM2008,Calgaro2015,Koide2018} as
\begin{equation}
\label{eqn_mvel}
U = U_v \,-\, \kam \,\nabla \ln \rho  = U_v \,-\, \frac{\kam}{\rho} \nabla \rho,
\end{equation}
where $\kam$ is the molecular diffusivity co-efficient. For simplicity, $\kam$ is assumed to be a constant.


\subsubsection{Re-casted continuity equation:}
It is straightforward to see that the classical continuity equation \eref{eqn_mass} transforms into a convection-diffusion type equation when the fluid mass velocity $U$ is replaced by the fluid volume velocity $U_v$ by using relation given by \eref{eqn_mvel}:
\begin{equation}
\label{eqn_rmass1}
\frac{\partial \rho}{\partial t} \, + \, \nabla \cdot [\rho \, U_v \,- \, \kam \,\nabla \rho ]  = 0.
\end{equation}
Assuming $\kam$ to be a constant, the re-casted mass balance equation have the following form:
\begin{equation}
\label{eqn_rmass2}
\frac{\partial \rho}{\partial t}  + \, \nabla \cdot [\rho \, U_v]  = \kam \, \Delta \rho,
\end{equation}
where $\Delta$ denotes the Laplacian operator.

\subsubsection{Re-casted momentum balance equation:}

We recast the full classical momentum balance equation \eref{eqn_momentum} by directly substituting \eref{eqn_mvel} into it. The full derivation is given in ~\ref{Appendix_B}. The final obtained expression of the momentum balance equation is given by:
\begin{eqnarray}
\label{eqn_rmomentum1}
\fl  \frac{\partial }{\partial t}  \left[ \rho \, U_v \,-\,\kam\, \nabla \rho \right]\,+ \,\nabla \cdot \Big[ \rho \,U_v \otimes U_v \,+\,  p\, \boldsymbol{I}\, +\, \bPi_v \, + \frac{\kam^2}{\rho} \nabla \rho \otimes \nabla \rho - \kam U_v \otimes \nabla \rho \nonumber \\
 \qquad \qquad \qquad \qquad \qquad \qquad - \kam  \nabla \rho \otimes U_v \Big] = 0.
\end{eqnarray}
The above re-casted momentum balance equation can be written with the help of the re-casted continuity equation \eref{eqn_rmass2} as:
\begin{eqnarray}
\label{eqn_rmomentum2}
\fl \frac{\partial \rho \, U_v }{\partial t} \, +\, \nabla \cdot \Big[ \rho \,U_v \otimes U_v \,+\,  p\, \boldsymbol{I}\, +\, \bPi_v \,+\, \frac{\kam^2}{\rho} \,\nabla \rho \otimes \nabla \rho -\, \kam\, U_v \otimes \nabla \rho \, -\, \kam \, \nabla \rho \otimes U_v \Big] \,\nonumber \\
\qquad \quad -\, \kam^2 \, \nabla \Delta \rho \,+\, \kam \,\nabla \left( \nabla \cdot (\rho \, U_v ) \right) = 0.
\end{eqnarray}
From \eref{eqn_rmomentum1}, one can extract a new stress tensor $\bPi^{(RNS)}_v$ as:
\begin{eqnarray}
\label{eqn_rstress1}
&& \bPi^{(RNS)}_v = \bPi_v\,+\, \frac{\kam^2}{\rho} \nabla \rho \otimes \nabla \rho \, - \,\kam \,U_v \otimes \nabla \rho \,-\, \kam \, \nabla \rho \otimes U_v,
\end{eqnarray}
where $\bPi_v$ denotes the transformed stress tensor from the classical stress tensor $\bPi^{(NS)}$ and is given by
\begin{eqnarray}
&& \bPi^{(NS)}\,\rightarrow\,\bPi_v = - 2 \mu \mathring{\overline{\nabla U_v}} \,+\,  2\, \mu \, \kam \,\widetilde{\bD} \ln \rho\, -\, \frac{2 \mu}{3} \kam \, \Delta \ln \rho\, \bI,
\end{eqnarray}
where $\widetilde{\bD}$ denotes the Hessian operator and its definition is given in \eref{eqn_A1}. Hence, another simplified form of the re-casted momentum balance may be given from equation \eref{eqn_rmomentum2} as:
\begin{eqnarray}
\label{eqn_rmomentum3}
\frac{\partial \rho \, U_v }{\partial t}\,+\, \nabla \cdot \left[ \rho \,U_v \otimes U_v\right]\,+\, \nabla \cdot \left[  p \, \boldsymbol{I} \,+\, \ \bPi^{(RNS)}_v \right]\,+ \, \kam \, \nabla \left[ \nabla \cdot (\rho \, U_v ) \right]\,\nonumber \\
\qquad \qquad \qquad \qquad \qquad \qquad \qquad \qquad \qquad \qquad -\, \kam^2  \nabla \Delta \rho \, = 0.
\end{eqnarray}
Using the identities listed in~\ref{Appendix_A} and later rearranging the terms in \eref{eqn_rstress1}, the expression for the new stress tensor $\bPi^{(RNS)}_v$ becomes:
\begin{eqnarray}
\label{eqn_rstress2}
\fl \bPi^{(RNS)}_v = \left( \frac{2}{3} \frac{\kam \,\mu}{\rho^2} |\nabla \rho|^2 \,-\, \frac{2}{3} \frac{\kam \mu}{\rho} \Delta \rho \right) \boldsymbol{I} \,-\,2\,\frac{\kam\,\mu}{\rho^2} \nabla \rho \otimes \nabla \rho \,-\,2\,\mu \,\mathbf{D}\left(U_v\right) \, \nonumber \\
\fl \qquad \qquad + \frac{2\,\mu}{3} \left( \nabla \cdot U_v\right) \boldsymbol{I} + \frac{\kam^2}{\rho} \nabla \rho \otimes \nabla \rho + 2\,\frac{\kam \,\mu}{\rho} \widetilde{\bD} \rho - \kam U_v \otimes \nabla \rho  - \kam  \nabla \rho \otimes U_v.
\end{eqnarray}
In order to compare the transformed stress tensor with the Korteweg stress tensor, it is convenient to write the re-casted momentum balance equation \eref{eqn_rmomentum2} in the following form:
\begin{eqnarray}
\label{eqn_rmomentum3}
&&  \frac{\partial \rho \, U_v }{\partial t}  + \nabla \cdot \left[ \rho \,U_v \otimes U_v \right]  =  \nabla \cdot \mbfT^{(RNS)} + \kam^2  \nabla \Delta \rho - \kam \nabla \left( \nabla \cdot (\rho \, U_v ) \right),
\end{eqnarray}
where $\mbfT^{(RNS)}$ is negative of the full pressure tensor and is given by:
\begin{eqnarray}
\label{eqn_rstress4}
\fl \mbfT^{(RNS)} = \left( - p - \frac{2}{3} \frac{\kam \,\mu}{\rho^2} |\nabla \rho|^2 + \frac{2}{3} \frac{\kam\,\mu}{\rho} \Delta \rho \right) \boldsymbol{I} + 2\,\frac{\kam\,\mu}{\rho^2} \nabla \rho \otimes \nabla \rho + 2\,\mu \,\mathbf{D}\left(U_v\right)\nonumber \\
\fl \qquad \qquad - \frac{2\,\mu}{3} \left( \nabla \cdot U_v\right) \boldsymbol{I} - \frac{\kam^2}{\rho} \nabla \rho \otimes \nabla \rho - 2\,\frac{\kam \,\mu}{\rho} \widetilde{\bD} \rho + \kam\, U_v \otimes \nabla \rho + \kam  \, \nabla \rho \otimes U_v.
\end{eqnarray}
The new stress tensor $\mbfT^{(RNS)}$ can be compared with the Korteweg stress tensor, $\mbfT$, proposed by Korteweg in $1901$ \cite{Korteweg1901}.
The Korteweg stress is dependant on the gradient of density in addition to the classical stress tensor, which is in turn dependent on the gradient of the velocity field alone, and is given by (see, equation (1.1) of \cite{Heida2010}):
\begin{eqnarray}
\fl \mbfT = \left( - p \,+\, \alpha_0 \,|\nabla \rho|^2\, +\, \alpha_1\, \Delta \rho \right) \boldsymbol{I} \,+\, \beta \left( \nabla \rho \otimes \nabla \rho \right) \,+\, 2\, \mu\, \mathbf{D}(\mathbf{v}) \,+\, \lambda (\nabla \cdot \mathbf{v}) \boldsymbol{I}.
\end{eqnarray}
Here, $p$ denotes the thermodynamic pressure, $\mathbf{D}(\mathbf{v})$ is the symmetric part of the velocity gradient and $\alpha_0, \alpha_1, \beta, \mu$ and $\lambda$ are material moduli that may depend on $\rho$ as well \cite{Heida2010}. It is note worthy to point out here that all terms involved in the Korteweg tensor $\mbfT$ are found in $\mbfT^{(RNS)}$ which is obtained by just re-casting the classical Navier-Stokes momentum balance equation in terms of the fluid volume velocity $U_v$.

For completeness, we also present the re-casted momentum balance equation in its non-conservative form
(see~\ref{Appendix_C} for the detailed derivation) and it is given by:
\begin{eqnarray}
&& \rho  \frac{\partial U_v }{\partial t}  \,+\, \rho \,(U_v \cdot \nabla) U_v \,+\, \nabla \cdot \left[ p \,\boldsymbol{I}\, + \,\bPi_v \,+\, \kam^2\, \left( \nabla \rho \otimes \frac{\nabla \rho}{\rho} \right) \right] \nonumber \\
&& \qquad \qquad - \kam \left[ \left(\nabla U_v - \widetilde{\nabla U_v}\right) \cdot \nabla \rho - \rho \nabla \left(\nabla \cdot U_v \right)  \right] - \kam^2  \nabla \Delta \rho = 0.
\end{eqnarray}
\subsubsection{Re-casted energy balance equation:}
The classical energy balance equation \eref{eqn_energy} can be re-casted in terms of the fluid volume velocity.  The detailed derivation is presented in ~\ref{Appendix_D}. The final form of the re-casted energy balance equation is
\begin{eqnarray}
\label{eqn_renergy}
\fl \frac{\partial}{\partial t} \left[\frac{1}{2} \rho U_v^2 + \rho \, \boldsymbol{e}_{in}\right] + \nabla \cdot \left[ \frac{1}{2} \rho \,U_v^2 \,U_v + \rho\, \boldsymbol{e}_{in}\, U_v \right] + \nabla \cdot \Bigg[ \left( p\,\boldsymbol{I} + \bPi_v \right) \cdot U_v - \kam \,\bPi_v \cdot \nabla \ln \rho \Bigg]\nonumber \\
\fl \qquad +\,\nabla \cdot \left[ \boldsymbol{q}^{(NS)} - \kam\,\Big( \rho \, \boldsymbol{e}_{in}\, \nabla \ln \rho + p\,\boldsymbol{I} \cdot \nabla \ln \rho \Big) \right]  + \nabla \cdot \Bigg[\kam \,\mcn_{v_1} + \kam^2\, \mcn_{v_2} + \kam^3\,\mcn_{v_3} \Bigg] \nonumber \\
\fl \qquad +\, \kam\, \mcn_{v_4} + \kam^2\, \mcn_{v_5} + \kam^3\, \mcn_{v_6} = 0,
\end{eqnarray}
where
\begin{eqnarray}
&& \mcn_{v_1} = -\,(U_v \cdot \nabla \rho)\, U_v \,-\,  \frac{1}{2}\, U_v^2 \,\nabla \rho, \\
&& \mcn_{v_2} = (U_v \cdot \nabla \rho) \, \nabla \ln \rho \,+\, \frac{1}{2\,\rho}\, |\nabla \rho|^2 \, U_v, \\
&& \mcn_{v_3} = - \frac{1}{2\, \rho}  |\nabla \rho|^2 \, \nabla \ln \rho,\\
&& \mcn_{v_4} = \nabla \cdot \left[ \rho \,U_v \otimes U_v + p \, \boldsymbol{I} + \ \bPi^{(RNS)}_v \right] \cdot \nabla \ln \rho \nonumber \\
&& \qquad \quad - U_v \cdot \Big[ \nabla \ln \rho\,\nabla \cdot (\rho\, U_v) - \nabla \left(\nabla \cdot (\rho\, U_v) \right) \Big], \\
&& \mcn_{v_5} = \Delta \rho\, \left( U_v \cdot  \nabla \ln \rho \right)\, -\, \left( U_v \cdot \nabla \Delta \rho \right) \,+\, \frac{1}{2} \frac{|\nabla\rho|^2}{\rho^2} \nabla \cdot \left(\rho\, U_v \right), \\
&& \mcn_{v_6} = -\,\frac{1}{2\,\rho^2} |\nabla\rho|^2\, \Delta \rho.
\end{eqnarray}
From the above energy balance equation, it is customary to observe that an expression for the new heat flux is given by:
\begin{equation}
\label{eqn-rhf1}
 \boldsymbol{q}^{(RNS)}_v = \boldsymbol{q}^{(NS)} \,- \,\kam\, \rho \,\boldsymbol{e}_{in}\, \nabla \ln \rho\,-\, \kam\,p\,\boldsymbol{I} \cdot \nabla \ln \rho.
\end{equation}
This final expression for the new heat flux gets also the following form:
\begin{equation}
\label{eqn-rhf2}
 \boldsymbol{q}^{(RNS)}_v =  - \kappa\, \nabla T \,- \,\kam\,\frac{\gamma}{(\gamma - 1)}\,p\,\nabla  \ln \rho.
\end{equation}

\subsubsection{Re-casted Navier-Stokes equations - I (RNS - I)}
The final set of the first re-casted Navier-Stokes equations obtained from the conventional Navier-Stokes by applying the velocity field transformation, $U = U_v - \kam \nabla \ln \rho$, is summarised as follows:
\begin{eqnarray}
\fl && \frac{\partial \rho}{\partial t}  + \, \nabla \cdot [\rho U_v]  - \kam \Delta \rho = 0, \label{eqn_RNSmass}\\
\fl && \frac{\partial \rho \, U_v }{\partial t}  + \nabla \cdot \left[ \rho \,U_v \otimes U_v +  p\, \boldsymbol{I} + \bPi^{(RNS)}_v \right] - \kam^2  \nabla \Delta \rho + \kam \nabla \left[ \nabla \cdot (\rho \, U_v ) \right] = 0, \label{eqn_RNSmoment} \\
\fl && \frac{\partial}{\partial t} \left[\frac{1}{2} \rho U_v^2 + \rho \, \boldsymbol{e}_{in}\right] \,+\,\nabla \cdot \left[ \frac{1}{2} \rho \,U_v^2 \,U_v \,+ \,\rho\, \boldsymbol{e}_{in}\, U_v \right]\,\nonumber \\
\fl && \qquad +\,\nabla \cdot \Big[ \left( p\,\boldsymbol{I}\, +\,\bPi_v \right) \cdot U_v\,- \, \kam \,\bPi_v \cdot \nabla \ln \rho \Big]\, +\,\nabla \cdot \left[\boldsymbol{q}^{(RNS)}_v\right] \nonumber \\
\fl && \qquad +\nabla \cdot \Big[\kam \,\mcn_{v_1} +\kam^2\, \mcn_{v_2} +\kam^3\,\mcn_{v_3} \Big]+\, \kam\, \mcn_{v_4}+\kam^2\, \mcn_{v_5}+ \kam^3\, \mcn_{v_6} = 0.\label{eqn_RNSenergy}
\end{eqnarray}

The continuum flow system \eref{eqn_RNSmass} - \eref{eqn_RNSenergy} is a type of mass diffusion set of continuum equations.
That is, it contains: (i) a mass diffusion component in the conservation of mass equation, (ii) explicit fluid dialation terms in the momentum stress tensor, and (iii) a non-Fourier heat flux term. The form of continuum flow equations \eref{eqn_RNSmass} - \eref{eqn_RNSenergy} appears more appropriate for flows involving large density variations/gradients. For example the Korteweg type shear stress components in \eref{eqn_RNSmoment} are found responsible for a better prediction of gas mass flow in a microchannel \cite{CD2018}.

\subsection{Re-casted Navier-Stokes equations - II (RNS - II): $U \rightarrow U_T - \kat \nabla \ln T$}
\label{2.3}

We derive a second set of re-casted Navier-Stokes by assuming that the flow mean mass velocity field $U$ can be written in terms of a new velocity field, which we call thermal diffusion velocity, $U_T$.
These two velocity fields are related by the following relation:
\begin{equation}
\label{eqn_tvel}
U = U_T - \kat \nabla \ln T = U_T - \frac{\kat}{T} \nabla T,
\end{equation}
where $\kat$ is the molecular thermal diffusivity co-efficient. Again, for simplicity, we assume $\kat$ to be a constant.

Following the same procedure used to derive the first re-casted Navier-Stokes (RNS - I) given in \S\ref{2.2}, we arrived at another set of equations from the classical equations and using the change of variable as defined in equation \eref{eqn_tvel}. We name these second re-casted equations, re-casted Navier-Stokes - II (RNS-II) and they are given by:
\begin{eqnarray}
\fl && \frac{\partial \rho}{\partial t}  + \, \nabla \cdot [\rho U_T] = \kat\, \rho\, \Delta \ln T \,+\,\kat\, \left(\nabla \rho \cdot \nabla \ln T \right) , \label{eqn_TRNSmass}  \\
\fl && \frac{\partial }{\partial t} \left( \rho \, U_T\,-\, \kat\, \rho\, \nabla \ln T \right)\,+\,\nabla \cdot \left[ \rho \,\left(U_T \otimes U_T \right) \right] \,+\,\nabla \cdot \left[ p\, \boldsymbol{I}\, +\,\bPi^{(RNS)}_T \right] = 0,\label{eqn_TRNSmomentum} \\
\fl && \frac{\partial}{\partial t} \left[\frac{1}{2} \rho\, U_T^2 + \rho\, \boldsymbol{e}_{in}\right]\,+\, \nabla \cdot \left[ \frac{1}{2} \rho\, U_T^2\, U_T\, +\, \rho \,\boldsymbol{e}_{in}\, U_T \right] \,\nonumber \\
\fl && \qquad \qquad +\,\nabla \cdot \left[(p \,\pmb{I} \,+\, \bPi_T) \cdot U_T\,-\,\kat\,\bPi_T \cdot  \nabla \ln T \right] \,+\, \nabla \cdot \left[ \boldsymbol{q}^{(RNS)}_T \right]\, \nonumber \\
\fl &&\qquad  \qquad +\,\nabla \cdot \Big[\kat \,\mcn_{T_1} \,+\,\kat^2\, \mcn_{T_2} \,+\,\kat^3\,\mcn_{T_3} \Big]\,+\,\kat\, \mcn_{T_4}\,+\,\kat^2\, \mcn_{T_5} = 0,\label{eqn_TRNSenergy}
\end{eqnarray}
where $\bPi^{(RNS)}_T,  \boldsymbol{q}^{(RNS)}_T$ and $\bPi_T$ are the new stress tensor, new heat-flux vector and the transformed classical stress tensor, respectively, and they are given by:
\begin{eqnarray}
\fl \bPi^{(RNS)}_T &=& \bPi_T\,- \rho\, \kat \, \left( U_T \otimes \nabla \ln T  + \nabla \ln T \otimes U_T \right) + \rho \,\kat^2 \nabla \ln T \otimes \nabla \ln T, \\
\fl \bPi_T &=&  - 2\, \mu \,\mathring{\overline{\nabla U_T}} \,+ \, 2 \,\mu \, \kat \, \widetilde{D} \ln T \, - \, \frac{2 \, \mu}{3} \,\kat \, \Delta \ln T \, \boldsymbol{I},\\
\fl \boldsymbol{q}^{(RNS)}_T &=&   \boldsymbol{q}^{(NS)} - \kat \, \rho \, \boldsymbol{e}_{in} \nabla \ln T - \kat \,p\, \pmb{I} \cdot \nabla \ln T,
\end{eqnarray}
and $\mcn_{T_i}$, $i={1,2,3,4,5}$ are additional terms which are given by:
\begin{eqnarray}
\fl \mcn_{T_1} &=& - \frac{1}{2}\rho \,U_T^2\, \nabla \ln T  \,-\, \rho\, (U_T \cdot \nabla \ln T)\, U_T, \\
\fl \mcn_{T_2} &=&  \rho\, (U_T \cdot \nabla \ln T) \, \nabla \ln T \,+\, \frac{1}{2} |\nabla \ln T|^2 \, \rho\,U_T,\\
\fl \mcn_{T_3} &=& - \frac{1}{2}  |\nabla \ln T|^2 \, \rho\,\nabla \ln T,\\
\fl \mcn_{T_4} &=& \nabla \cdot \left[ \rho \,\left(U_T \otimes U_T \right)\,+\, p\, \boldsymbol{I}\, +\,\bPi^{(RNS)}_T \right]\,\cdot \nabla \ln T \,-\,\rho\,U_T \cdot  \nabla \left(\frac{1}{T} \frac{\partial T}{\partial t} \right), \\
\fl \mcn_{T_5} &=& -\,\frac{1}{2}\,|\nabla \ln T|^2\, \left( \frac{\partial \rho}{\partial t} \right).
\end{eqnarray}

The continuum flow model \eref{eqn_TRNSmass} - \eref{eqn_TRNSenergy} is a mass diffusion continuum flow model similar to RNS - I. The diffusion component in the mass conservation equation is now driven by the temperature gradient. The temperature gradient term contributions to its momentum balance shear stress tensor are similar to the Ghost effect term contributions to the classical Navier-Stokes claimed by \cite{Sone1966,Sone2000,Sone2003,Sone2004}. Those terms are responsible for the predictions of, for example, thermophoresis and other thermal stress driving flows \cite{Sone1966,Sone2000,Sone2003,Sone2004}.

\subsection{Re-casted Navier-Stokes equations - III (RNS - III): $U \rightarrow U_p - \kap  \nabla \ln p$}
\label{2.4}
In this subsection, we recast the classical Navier-Stokes equations using a new relation in which the flow mean mass velocity field $U$ is related to the new velocity field called pressure diffusion velocity, $U_p$, by:
\begin{equation}
\label{eqn_pvel}
U = U_p \,- \,\kap \, \nabla \ln p\, =\, U_p\, -\, \frac{\kap}{p} \nabla p,
\end{equation}
where $\kap$ is the molecular pressure diffusivity co-efficient and, for simplicity, is assumed to be a constant.

We name the re-casted Navier-Stokes derived from equation \eref{eqn_pvel}, re-casted Navier-Stokes - III (RNS - III) and their derivation procedure is the same as that followed to obtain RNS - I and RNS - II. The re-casted Navier-Stokes - III are then given by:
\begin{eqnarray}
\fl && \frac{\partial \rho}{\partial t}  + \, \nabla \cdot [\rho\,U_p] = \kap\,\rho\, \Delta \ln p \,+\,\kap\, \left(\nabla \rho \cdot \nabla \ln p \right), \label{eqn_PRNSmass}  \\
\fl && \frac{\partial }{\partial t} \left( \rho \, U_p\,-\, \kap\, \rho\, \nabla \ln p \right)\,+\,\nabla \cdot \left[ \rho \,\left(U_p \otimes U_p \right) \right] \,+\,\nabla \cdot \left[ p\, \boldsymbol{I}\, +\,\bPi^{(RNS)}_p \right] = 0,\label{eqn_PRNSmomentum}  \\
\fl && \frac{\partial}{\partial t} \left[\frac{1}{2} \rho\, U_p^2 + \rho\, \boldsymbol{e}_{in}\right]+ \nabla \cdot \left[ \frac{1}{2} \rho\, U_p^2\, U_p + \rho \,\boldsymbol{e}_{in}\, U_p \right] + \nabla \cdot \left[(p \,\pmb{I} + \bPi_p) \cdot U_p - \kap\,\bPi_p \cdot  \nabla \ln p \right] \,\nonumber \\
\fl && \qquad +\, \nabla \cdot \left[ \boldsymbol{q}^{(RNS)}_p \right]\, \,+\,\nabla \cdot \Big[\kap \,\mcn_{p_1} \,+\,\kap^2\, \mcn_{p_2} \,+\,\kap^3\,\mcn_{p_3} \Big]\,+\,\kap\, \mcn_{p_4}\,+\,\kap^2\, \mcn_{p_5} = 0, \label{eqn_PRNSenergy}
\end{eqnarray}
where $\bPi^{(RNS)}_p, \boldsymbol{q}^{(RNS)}_p$ and $\bPi_p$ are generalized stress tensor, generalized heat-flux vector and the transformed classical stress tensor, respectively and are given by:
\begin{eqnarray}
\fl \bPi^{(RNS)}_p &=& \bPi_p\,- \rho\, \kap \, \left( U_p \otimes \nabla \ln p  + \nabla \ln p \otimes U_p \right) + \rho \,\kap^2 \nabla \ln p \otimes \nabla \ln p, \\
\fl \bPi_p &=&  - 2\, \mu \,\mathring{\overline{\nabla U_p}} \,+ \, 2 \,\mu \, \kap \, \widetilde{\bD} \ln p \, + \, \lambda \,\kap \, \Delta \ln p \, \boldsymbol{I},\\
\fl \boldsymbol{q}^{(RNS)}_p &=&  \boldsymbol{q}^{(NS)} - \kap \, \rho \, \boldsymbol{e}_{in} \nabla \ln p - \kap \,p\, \pmb{I} \cdot \nabla \ln p,
\end{eqnarray}
and $\mcn_{p_i}$, $i={1,2,3,4,5}$ are additional terms whose expressions are given by:
\begin{eqnarray}
\fl \mcn_{p_1} &=&  - \frac{1}{2}\rho \,U_p^2\, \nabla \ln p  \,-\, \rho\, (U_p \cdot \nabla \ln p)\, U_p,\\
\fl \mcn_{p_2} &=& \rho\, (U_p \cdot \nabla \ln p) \, \nabla \ln p \,+\, \frac{1}{2} |\nabla \ln p|^2 \, \rho\,U_p,\\
\fl \mcn_{p_3} &=& - \frac{1}{2}  |\nabla \ln p|^2 \, \rho\,\nabla \ln p,\\
\fl \mcn_{p_4} &=&  \nabla \cdot \left[ \rho \,\left(U_p \otimes U_p \right)\,+\, p\, \boldsymbol{I}\, +\,\bPi^{(RNS)}_p \right]\,\cdot \nabla \ln p \,-\,\rho\,U_p \cdot \nabla \left(\frac{1}{p} \frac{\partial p}{\partial t} \right),\\
\fl \mcn_{p_5} &=& - \frac{1}{2}\,|\nabla \ln p|^2\, \left( \frac{\partial \rho}{\partial t} \right).
\end{eqnarray}

As opposed to the previous re-casted NS, the mass diffusion in \eref{eqn_PRNSmass} - \eref{eqn_PRNSenergy} is now driven by the pressure gradient. Hence, the system \eref{eqn_PRNSmass} - \eref{eqn_PRNSenergy} may therefore be applicable to liquid flows. Consider a two-dimensional isothermal pressure driven flow, one observes, that the additional contributions of the pressure gradient terms in the shear stress in equation \eref{eqn_PRNSmomentum} may lead to additional contributions to the flow rate. These diffusive terms may also be responsible for interpreting some of the high flows of water in nano tubes as already demonstrated in gas flows \cite{VeltzkeThAming2012}.

\section{Linear stability analysis and sound dispersion}
\label{3}
In this section we examine both temporal and spatial stability analyses of the three new sets of re-casted Navier-Stokes equations derived in \S \ref{2.2}, \S \ref{2.3} and \S \ref{2.4}. We consider the re-casted Navier-Stokes models in a one dimensional flow configuration.
An equilibrium ground state is defined by the flow variables $\rho_o$, $T_o$, $p_o= \bR \,\rho_o\, T_o$, $u_o = 0$, with $\bR$ as the specific gas constant. A perturbation to the equilibrium ground state is introduced as follows:
\begin{eqnarray}
\rho = \rho_o(1+\rho^*), \ T = T_o(1+T^*), \ u = u^* \sqrt{\bR T_o}, \  p = p_o(1+p^*),
\end{eqnarray}
where the asterisked variables represent dimensionless quantities with $p^*= \rho^*+T^*$, and the subscript $_o$ denotes the equilibrium ground state flow parameters. The dimensionless space and time variables are specified using a characteristic length $L$ and a characteristic time $\tau$ by the expressions:
\begin{eqnarray}
x = L x^*, \quad t= \tau\, t^*, \quad  \tau= \frac{L}{\sqrt{\bR T_o}}.
\end{eqnarray}
The corresponding dimensionless transport coefficients are given by:
\begin{eqnarray}
\fl && \mu^* = \frac{\mu}{L \,\rho_o \,\sqrt{\bR\, T_o}} = \frac{\mu}{\mu_o}, \quad \kam^* = \frac{\kam}{L \, \sqrt{\bR \, T_o}} = \frac{\kam \,\rho_o}{\mu_o}, \quad \alpha_p^* = \frac{\alpha_p}{L \, \sqrt{\bR \, T_o}} = \frac{\alpha_p \,\rho_o}{\mu_o},\nonumber \\
\fl &&   \kat^* = \frac{\kat}{L \,\sqrt{\bR \,T_o}} = \frac{\kat\, \rho_o}{\mu_o}, \quad \kappa^* = \frac{\kappa}{\bR \,L\, \rho_o \sqrt{\bR \,T_o}} =\frac{\kappa}{\bR\, \mu_o},
\end{eqnarray}
where $\mu_o$ is a reference viscosity coefficient chosen such that the Knudsen number, $\rm{Kn}$, is set equal to unity.

The dimensionless form of linearized re-casted Navier-Stokes equations - I (RNS - I) are given by:
\begin{equation}
\label{eqn_LRNS-I}
\eqalign{ \frac{\partial \rho^*}{\partial t^*}  + \frac{\partial u_v^*}{\partial x^*}  - \kam^*  \frac{\partial^2 \rho^*}{\partial x^{*^2}} = 0, \\
 \frac{\partial u_v^*}{\partial t^*} +\frac{\partial \rho^*}{\partial x^*} +\frac{\partial T^*}{\partial x^*} - \left(\frac{4}{3} \mu^* - \kam^* \right) \frac{\partial^2 u_v^*}{\partial x^{*^2}} + \left(\frac{4}{3} \mu^* \kam^* - \kam^{*^2} \right) \frac{\partial^3 \rho^*}{\partial  x^{*^3}} = 0, \\
 \frac{\partial T^*}{\partial t^*} +\frac{2}{3}\frac{\partial u_v^*}{\partial x^*}- \frac{2}{3} \kam^*  \frac{\partial^2 \rho^*}{\partial x^{*^2}}\,-\, \frac{2}{3} \kappa^*   \frac{\partial^2 T^*}{\partial x^{*^2}}  = 0.}
\end{equation}

The dimensionless form of linearized version of re-casted Navier-Stokes equations - II (RNS - II) are given by:
\begin{equation}
\label{eqn_LRNS-II}
\eqalign{ \frac{\partial \rho^*}{\partial t^*}\, +\, \frac{\partial u_T^*}{\partial x^*} \, - \,\kat^* \, \frac{\partial^2 T^*}{\partial x^{*^2}} = 0,\\
 \frac{\partial u_T^*}{\partial t^*}\, -\, \left(\frac{4}{3}\, \mu^* \,-\, \frac{2}{3} \,\kat^* \right) \frac{\partial^2 u_T^*}{\partial x^{*^2}}\,+\,\frac{\partial \rho^*}{\partial x^*}\, +\,\frac{\partial T^*}{\partial x^*}\,  \\
 \qquad \,\,+\, \left(\frac{4}{3}\, \mu^* \, -\, \frac{2}{3} \,\kat^* \,-\, \frac{2}{3}\, \kappa \right) \,\kat^* \,\frac{\partial^3 T^*}{\partial  x^{*^3}}\, = 0, \\
\frac{\partial T^*}{\partial t^*}\, +\,\frac{2}{3}\frac{\partial u_T^*}{\partial x^*}\,-\, \frac{2}{3} \left(\ka \,+\, \kat \right)   \frac{\partial^2 T^*}{\partial x^{*^2}}  = 0.}
\end{equation}

Finally, the dimensionless form of linearized re-casted Navier-Stokes equations - III (RNS - III) are given by:

\begin{equation}
\label{eqn_LRNS-III}
\eqalign{ \frac{\partial \rho^*}{\partial t^*}\,+\, \frac{\partial u_p^*}{\partial x^*}\,-\, \kap\,\frac{\partial^2 \rho^*}{\partial x^{*^2}} \,-\, \kap  \frac{\partial^2 T^*}{\partial x^{*^2}} = 0, \\
 \frac{\partial u_p^*}{\partial t^*}\,-\, \left(\frac{4}{3} \mu^* \,-\, \frac{5}{3} \kap \right) \frac{\partial^2 u_p^*}{\partial x^{*^2}}\, +\,\frac{\partial \rho^*}{\partial x^*} \,+\, \left(\frac{4}{3} \mu^* \,-\, \frac{5}{3} \kap \right) \, \kap \, \frac{\partial^3 \rho^*}{\partial x^{*^3}} \\
 \qquad  +\,\frac{\partial T^*}{\partial x^*}  \,+\, \left(\frac{4}{3} \mu^*\,-\, \frac{5}{3} \kap \,-\, \frac{2}{3} \ka \right) \kap \,\frac{\partial^3 T^*}{\partial  x^{*^3}} = 0, \\
  \frac{\partial T^*}{\partial t^*} +\frac{2}{3}\frac{\partial u_p^*}{\partial x^*}- \frac{2}{3} \kap  \frac{\partial^2 \rho^*}{\partial x^{*^2}}\,-\, \frac{2}{3} \left(\ka + \kap \right)   \frac{\partial^2 T^*}{\partial x^{*^2}}  = 0.}
\end{equation}

We assume the disturbances $\rho^*$, $T^*$ and $U_v^*$ to be wave functions of the form
\begin{equation}
\label{eqn_PWS}
\phi^* = \phi_a^* \, \exp [i\, (\omega\, t^* - {\rm K}\, x^*)],
\end {equation}
where $\omega$ is the complex wave frequency, $K$ is the complex wave number, and $\phi_a^*$ is the complex amplitude, so that we have
\begin{eqnarray}
 \frac{\partial \phi^*}{\partial t^*} = i \,\omega \,\phi^*, \quad \frac{\partial \phi^*}{\partial x^*} = - i \,{\rm K} \,\phi^*,  \quad \frac{\partial^2 \phi^*}{\partial x^{*^2}} = - {\rm K}^2 \,\phi^*,  \quad \frac{\partial^3 \phi^*}{\partial x^{*^3}} =  i \,{\rm K}^3 \,\phi^*.
\end{eqnarray}

Substitution of the plane wave solution \eref{eqn_PWS} into the linearized re-casted Navier-Stokes - I system given by \eref{eqn_LRNS-I} yields the homogeneous system $\sf{A}(\omega, {\rm K})\, \phi^* = 0$, where
\begin{equation}
\fl{ \sf{A}(\omega, {\rm K}) =
\left[\begin{array}{ccc}
i \omega + \kam^* {\rm K}^2 & 0 & -i {\rm K} \\
\frac{2}{3} \kam^* {\rm K}^2  & i \omega + \frac{2}{3}  \kappa^* {\rm K}^2 & -\frac{2}{3} i {\rm K} \\
-i {\rm K} + \left( \frac{4}{3} \mu^* \kam^* - \kam^{*^2} \right) i {\rm K}^3 & - i {\rm K}  & i \omega + \left(\frac{4}{3} \mu^* - \kam^* \right) {\rm K}^2
\end{array} \right] }
\end{equation}
and
\begin{equation}
\phi^* =
\left[\begin{array}{c}
\rho^* \\
T^* \\
U_v^*
\end{array} \right].
\end{equation}

%
%
The corresponding dispersion relation, obtained when the determinant of $\mathcal{A} (\omega, {\rm K})$ is zero, is
\begin{equation}
\label{eqn_dispersion}
 9 i \omega^3 + 6 {\rm K}^2 \left(\kappa^* + 2 \mu^*\right) \omega^2 - i {\rm K}^2 \left(8 {\rm K}^2 \kappa^* \mu^* + 15 \right) \omega - 6 {\rm K}^4  \kappa^* = 0.
\end{equation}

\begin{figure}
\centering
\includegraphics[height=6.5cm]{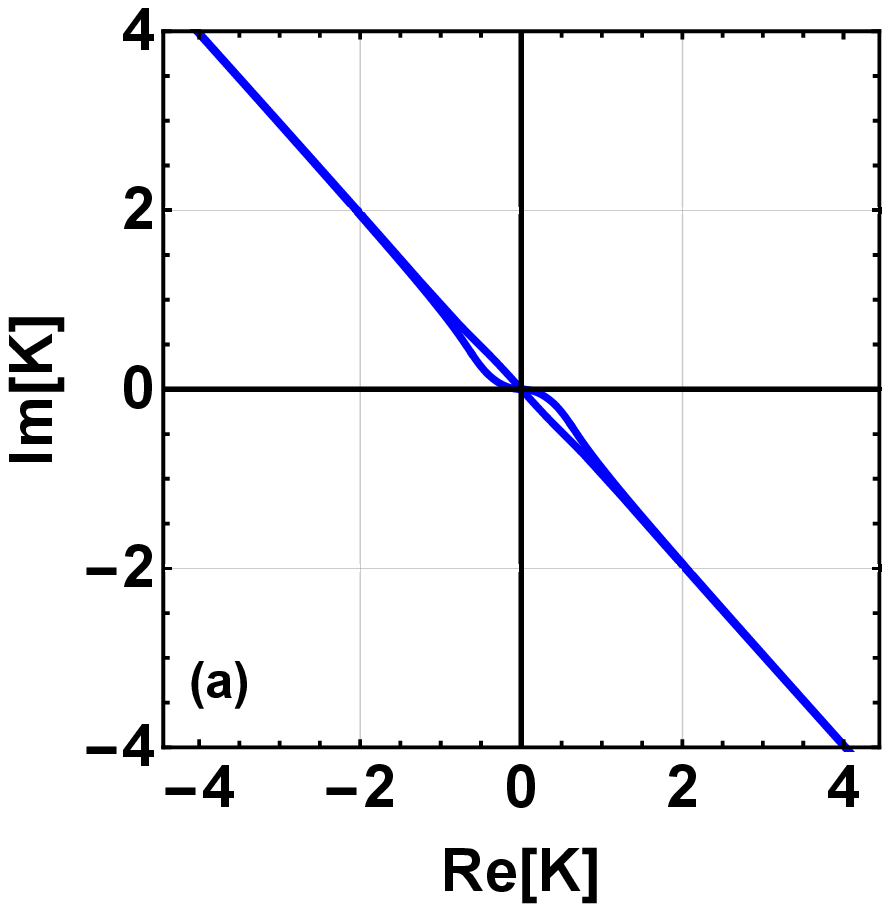}\quad
\includegraphics[height=6.5cm]{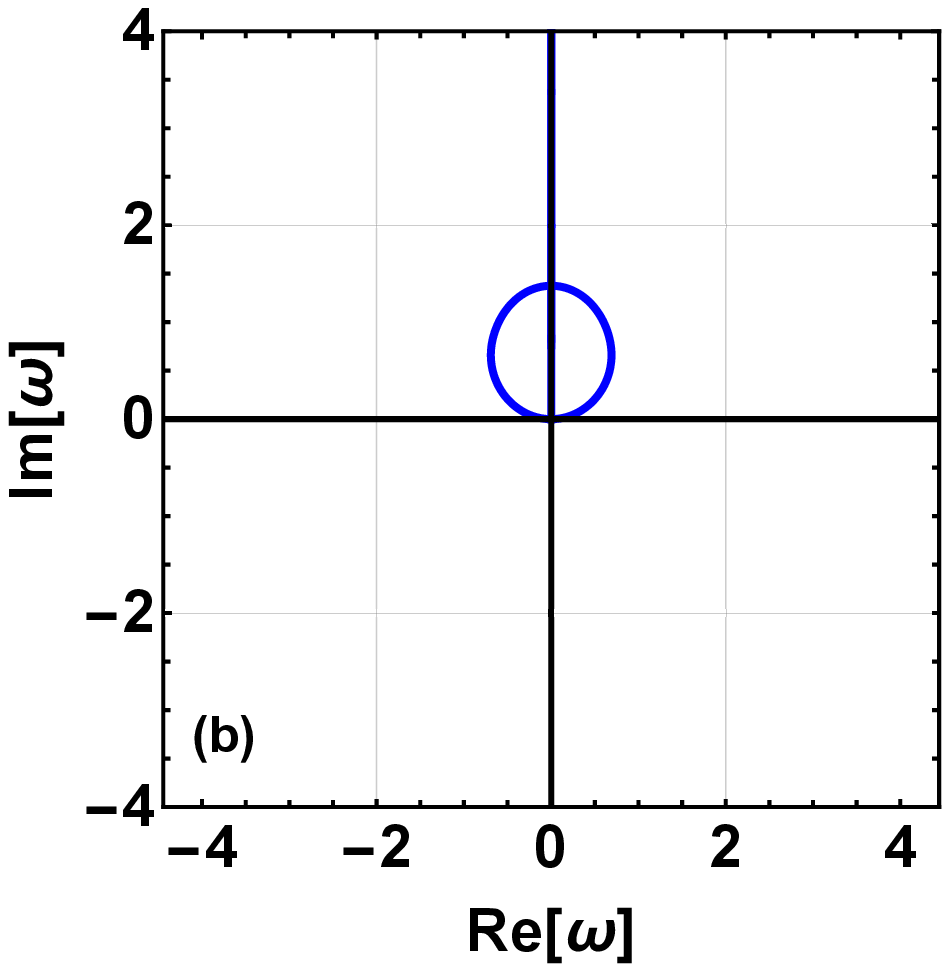}
\caption{Stability analysis of the re-casted Navier-Stokes equations: panel (a) spatial stability  and panel (b) temporal stability with $\kappa^* = 15/4$.}
\label{fig:1}
\end{figure}


\begin{figure}
\centering
\includegraphics[height=6.5cm]{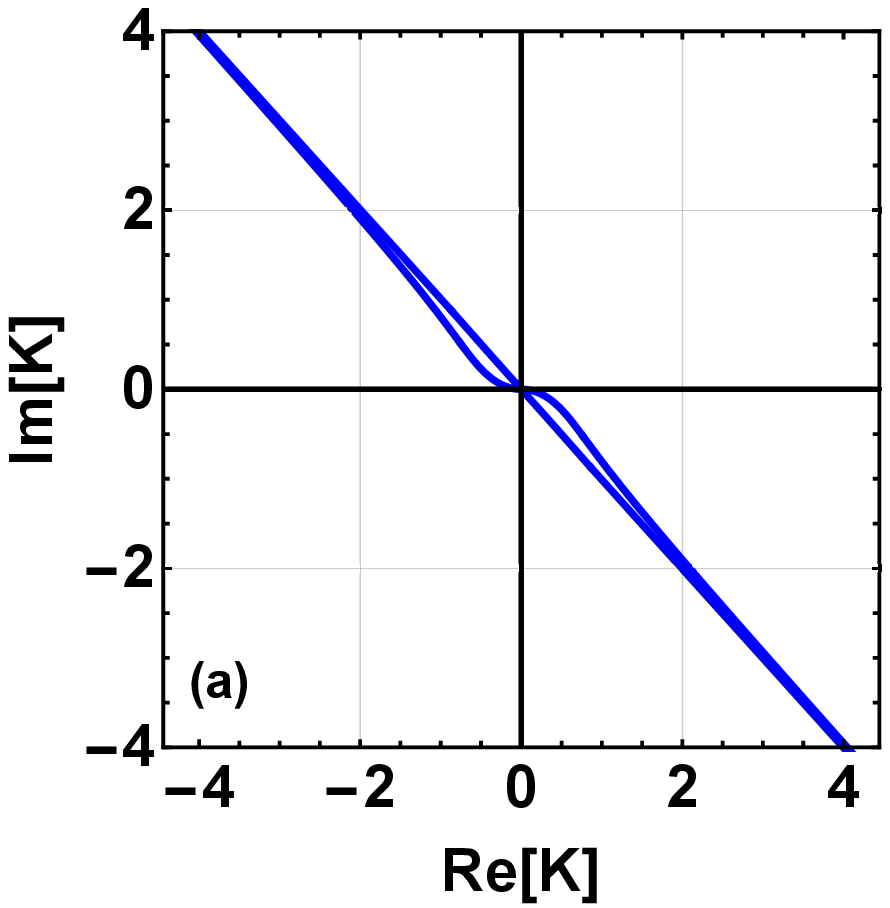}\quad
\includegraphics[height=6.5cm]{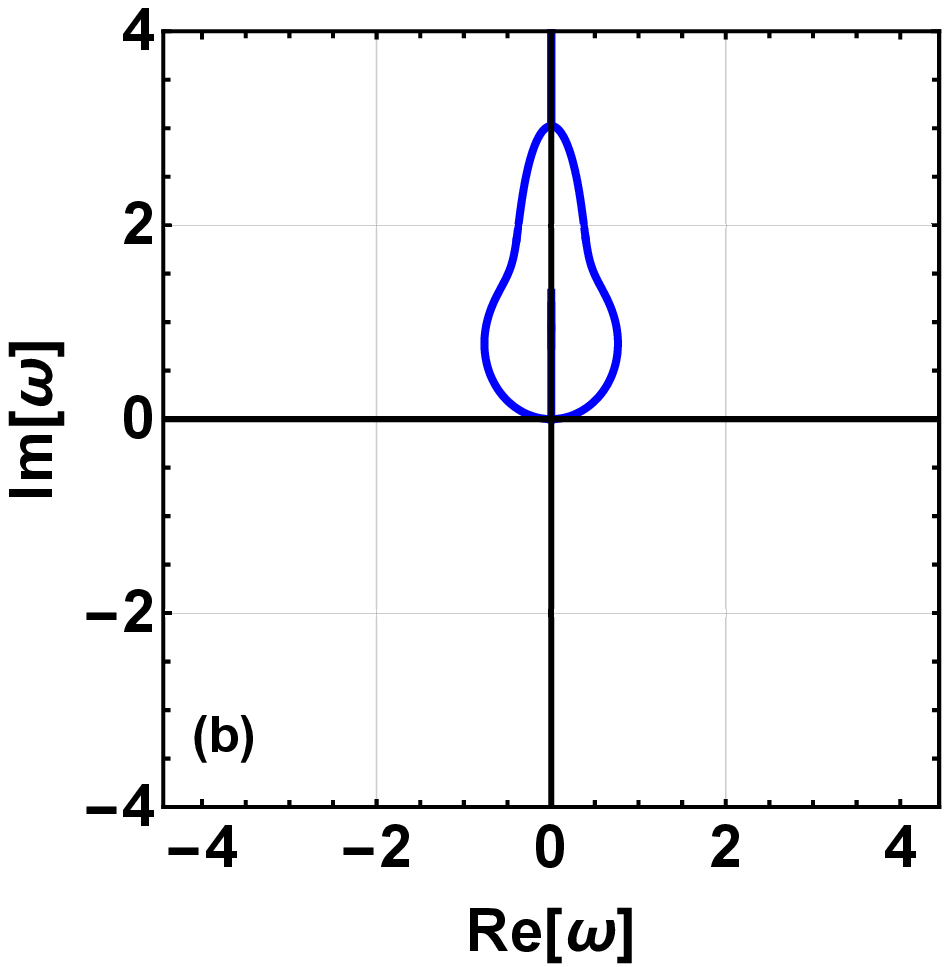}
\caption{Stability analysis of the re-casted Navier-Stokes equations: panel (a) spatial stability  and panel (b) temporal stability with $\kappa^* = 13/4$.}
\label{fig:2}
\end{figure}

\begin{figure}
\centering
\includegraphics[height=6.5cm]{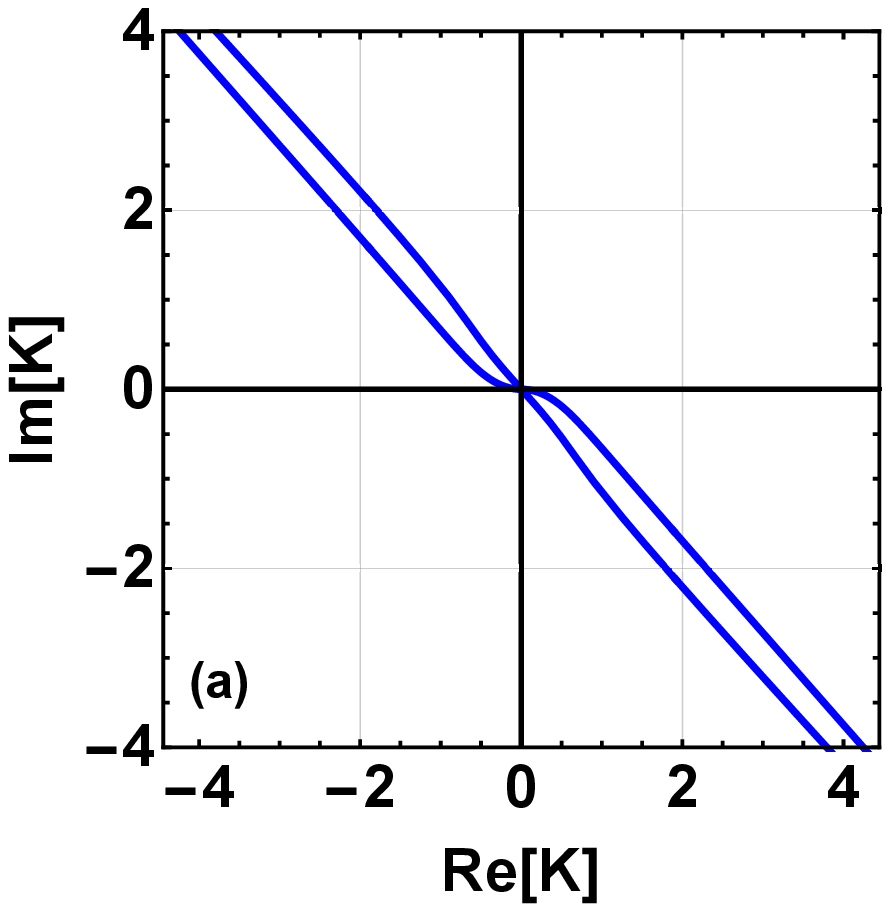}\quad
\includegraphics[height=6.5cm]{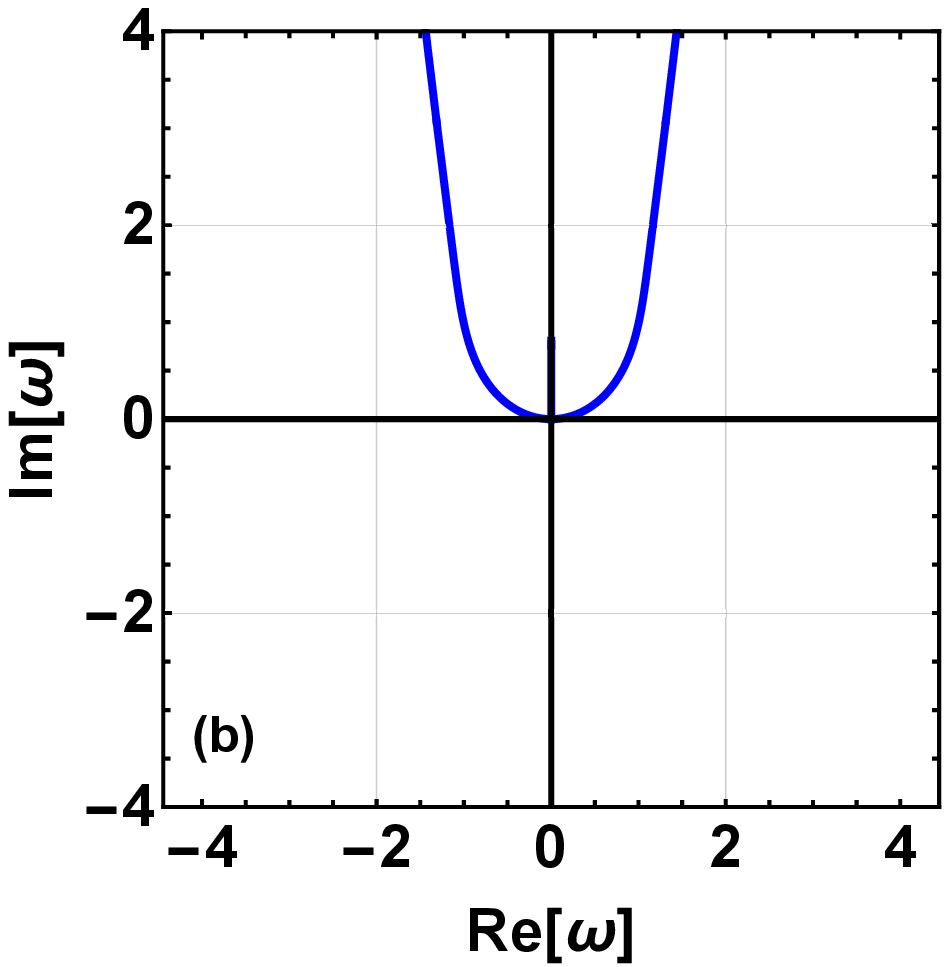}
\caption{Stability analysis of the re-casted Navier-Stokes equations: panel (a) spatial stability  and panel (b) temporal stability with $\kappa^* = 9/4$.}
\label{fig:3}
\end{figure}
We observe that dispersion relation \eref{eqn_dispersion} does not depend explicitly upon $\kam^*$, which is the dimensionless molecular diffusivity co-efficient. In fact, this dispersion relation is the same as that of the classical Navier-Stokes equations.
In \cite{Dadzie2013} it was shown that transport coefficients associated with mass/volume diffusion theory may be different from transport coefficients when the classical theory is in use. Hence, we examine the stability of the re-casted NS equations by considering different transport coefficients: (i) $\kappa^* = 15/4$ which corresponds to the value from the classical theory, (ii) $\kappa^* = 13/4$ which is an assumed value for the re-casted theory, and (iii) $\kappa^* = 9/4$ which corresponds to the value from the volume-diffusion theory of \cite{Dadzie2013}. Figure \ref{fig:1}, shows the spatial and temporal stability of the re-casted Navier-Stokes for $\kappa^* = 15/4$. For this value of $\kappa^*$, the stability diagram presented in figure \ref{fig:1} is exactly the same as that of the classical Navier-Stokes. Figure \ref{fig:2} and figure \ref{fig:3} present stability results for the re-casted Navier-Stokes equations for $\kappa^* = 13/4$ and $\kappa^* = 9/4$, respectively. The new re-casted Navier-Stokes models are unconditionally stable in both space and time like the classical Navier-Stokes model for all these coefficients.
The same conclusions are reached for re-casted Navier-Stokes equations - II (RNS - II) and re-casted Navier-Stokes equations - III (RNS - III).
\begin{figure}
\centering
\includegraphics[height=6.5cm]{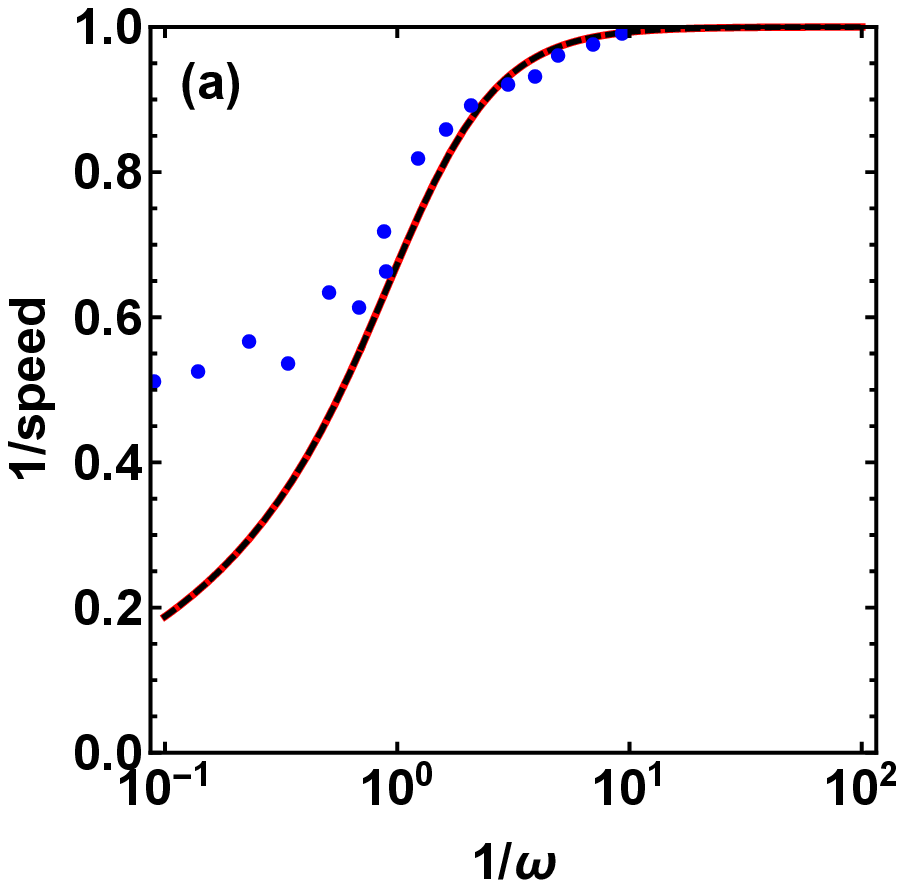}\quad
\includegraphics[height=6.5cm]{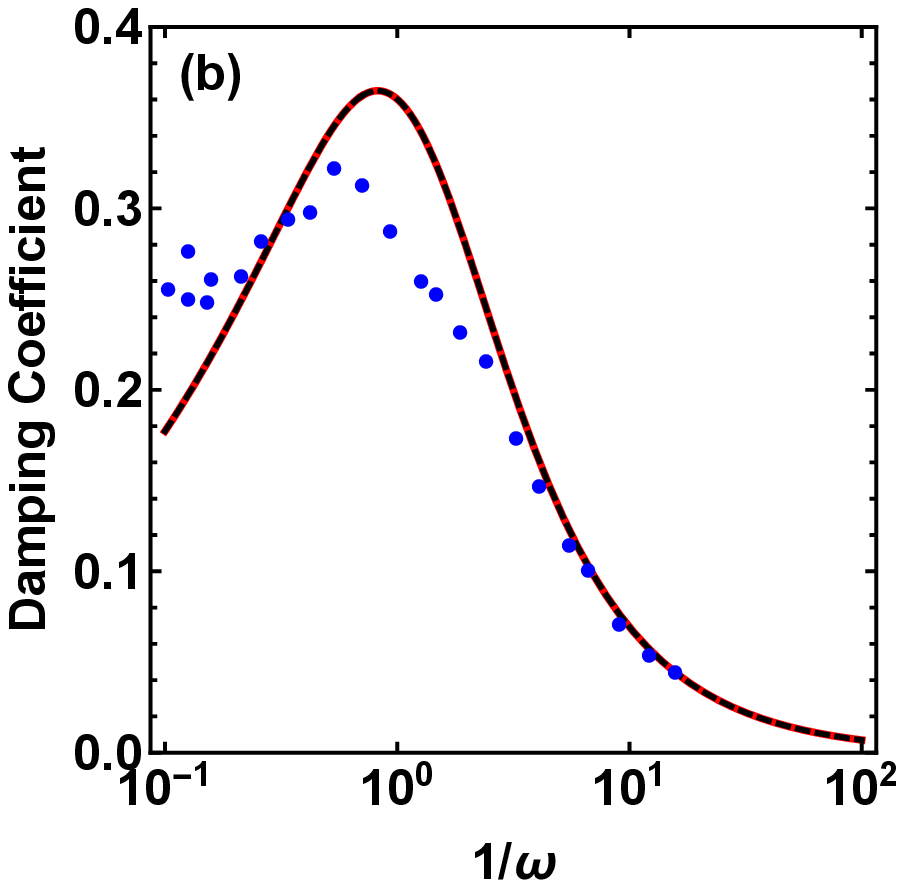}
\caption{Sound dispersion with $\kappa^* = 15/4$: panel (a) inverse sound speed and panel (b) damping coefficient. Solid and dotted lines represents the results from the re-casted NS model and the classical NS model, respectively. Filled circles represents the experimental results by Meyer and Sessler~\cite{Meyer1957}.}
\label{fig:4}
\end{figure}

\begin{figure}
\centering
\includegraphics[height=6.5cm]{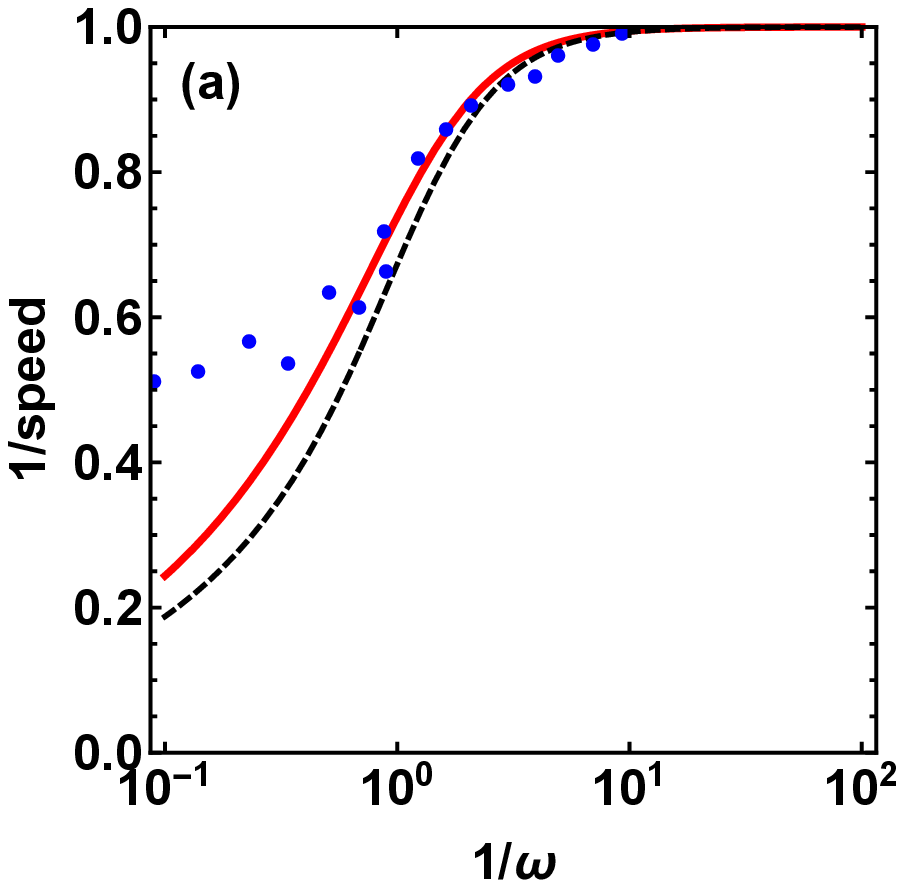}\quad
\includegraphics[height=6.5cm]{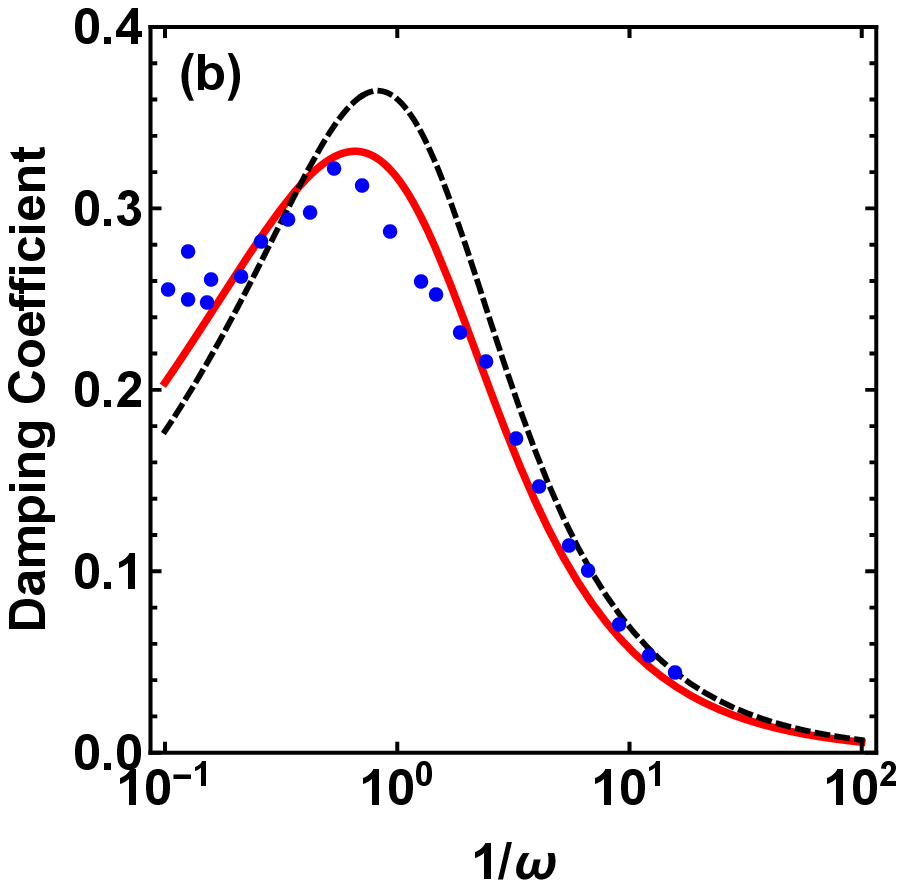}
\caption{Sound dispersion with $\kappa^* = 9/4$: panel (a) inverse sound speed and panel (b) damping coefficient. Solid and dotted lines represents the results from the re-casted NS model and the classical NS model, respectively. Filled circles represents the experimental results by Meyer and Sessler~\cite{Meyer1957}.}
\label{fig:5}
\end{figure}

We now analyze the re-casted model predictions of sound dispersions as compared with the experimental data from Meyer and Sessler~\cite{Meyer1957}. The dimensionless inverse of phase speed and dimensionless spatial damping coefficient are commonly defined as \cite{Greenspan1950,Meyer1957,Dellar2007,Dadzie2010}
\begin{equation}
\sqrt{\frac{5}{3}} \frac{\rm{Re}[K]}{\omega} \quad \mbox{and} \quad - \sqrt{\frac{5}{3}} \frac{\rm{Im}[K]}{\omega}.
\end{equation}
Predictions by the re-casted model with $\kappa^* = 15/4$ and $\kappa^* = 9/4$ for the inverse sound speed and damping coefficient are plotted in figure \ref{fig:4} and figure \ref{fig:5}, respectively, alongside predictions by classical Navier-Stokes and the experimental data by Meyer and Sessler~\cite{Meyer1957}. When $\kappa^* = 15/4$ predictions of the inverse sound speed and damping coefficients by the re-casted NS models exactly coincide with that of the classical NS as expected. With $\kappa^* = 9/4$, the re-casted NS model achieved a better agreement with the experimental data as compared to the classical Navier-Stokes model which is evident from figure \ref{fig:5}. This good agreement with the experimental data is the same as that obtained in the volume diffusion model of Dadzie \cite{Dadzie2013} using the same coefficient.

\section{Application to the Rayleigh-Brillouin light scattering experiment}
\label{4}
In order to clearly demonstrate the usefulness of the re-casted Navier-Stokes models derived in \S \ref{2.2}, \S \ref{2.3} and \S \ref{2.4}, we analyse the Rayleigh-Brillouin light scattering problem.
When light is passed through a gas, some parts of the light are scattered by the motion of the gas molecules \cite{Mountain1966}.
The spectral profile/distribution of the scattered light mainly depends on the scattering mechanism.
In particular, the Rayleigh-Brillouin scattering (RBS) combines two scattering mechanisms:
Rayleigh scattering and Brillouin scattering. Both originate when the light is scattered due to density fluctuations of gas molecules. An RBS spectrum consists of two dominant components: the Rayleigh part due to thermal motion of the gas molecules, which causes a Doppler shift relative to the incident wavelength, and the Brillouin part, which is due to the exchange of energy between light and acoustic modes in the medium and is associated with the acoustic effect of gas molecules \cite{Ma2014}. Rayleigh-Brillouin scattering is nowadays a powerful method to investigate thermodynamic properties of transparent media such as gas, water and optical fibers \cite{Witschas2010,Gu2012,Gu2013,Fry2002,Schorstein2009,Huang2012,Shi2012,Xie2012}. RBS is used to determine the physical properties of gases, such as sound speed, thermal diffusivity, heat capacity ratios, bulk viscosity, etc. \cite{Ma2014}.
It is also used in oceanographic studies to find out the ocean salinity, temperature, sound speed and viscosity \cite{Schorstein2008,Liang2011,Liang2012,Xu2003,Liu2002,Ma2014}.

\subsection{Formulation}
\label{4.1}

The spectrum of the scattered light follows from the knowledge of the gas density fluctuations (the density-density correlation function) and are obtained from the linearized hydrodynamic models \cite{Marques1993}. The gas density fluctuations can either arise spontaneously or are created by external optical potentials. Based on the gas density fluctuations, we have the spontaneous Rayleigh-Brillouin scattering (SRBS) or the coherent Rayleigh-Brillouin scattering (CRBS). The spectrum of the scattered light depends on the Knudsen number, which is related to a frequently used $y$ parameter in RBS experiments \cite{Marques1993,Marques1998,Wu2016}, intermolecular potentials and the rotational collision number.

In the spontaneous RBS, an incident light wave vector $\bk_i$ is scattered with scattered-light wave vector $\bk_s$ due to the spontaneous density variations of gas molecules. If the scattering angle is $\theta$ then the scattering wave number ($k$) is given by:
\begin{equation}
k = |\bk_i - \bk_s| = 2\, |\bk_i|\, \sin(\theta/2).
\end{equation}
In the coherent RBS, light is scattered through the density fluctuations of gas molecules which are generated from the interference pattern induced by two plumb laser beams \cite{Pan2002,Pan2004}. For the physical process and the experimental set-up of spontaneous RBS and coherent RBS, readers are referred to references \cite{Pan2002,Pan2004}.

The one-dimensional, linearized nature and lack of boundary conditions make the hydrodynamic equations simple and allow for analytical solutions for the Rayleigh-Brillouin scattering problem \cite{Ma2012,Marques1993,Wu2018}. The spectrum of the scattered light is characterized by the Knudsen number ($\rm{Kn}$), which is here the ratio of the mean free path of gas molecules to the characteristic length scale ($L$) of the system, identified as the scattering wave length $2\,\pi/k$. We use the same dimensionless variables which are in line with the previous linear analysis (\S \ref{3}) and one additional parameter, namely, the Knudsen number to linearize the re-casted Navier-Stokes models derived in \S \ref{2.2}, \S \ref{2.3} and \S \ref{2.4}.

The linearized form of re-casted Navier-Stokes equations - I is:
\begin{equation}
\label{eqn_RBSLRNS-I}
\fl \eqalign{ \frac{\partial \rho^*}{\partial t^*}  + \frac{\partial u_v^*}{\partial x^*}  - \kam^* \,{\rm{Kn}} \frac{\partial^2 \rho^*}{\partial x^{*^2}} = 0, \\
\frac{\partial u_v^*}{\partial t^*} +\frac{\partial \rho^*}{\partial x^*} +\frac{\partial T^*}{\partial x^*} - \left(\frac{4}{3} - \kam^* \right) \,{\rm{Kn}}\, \frac{\partial^2 u_v^*}{\partial x^{*^2}} + \left(\frac{4}{3} \kam^* - \kam^{*^2} \right) \,{\rm{Kn}^2}\, \frac{\partial^3 \rho^*}{\partial  x^{*^3}} = 0, \\
 \frac{\partial T^*}{\partial t^*} +\frac{2}{3}\frac{\partial u_v^*}{\partial x^*}- \frac{2}{3} \kam^*  \,{\rm{Kn}} \frac{\partial^2 \rho^*}{\partial x^{*^2}}\,-\, \frac{2}{3} \kappa^* \,{\rm{Kn}}   \frac{\partial^2 T^*}{\partial x^{*^2}} = 0.}
\end{equation}

In order to obtain the spontaneous RBS, one needs to apply the temporal Laplace transform and the spatial Fourier transform of the density fluctuations. Hence, the spontaneous spectrum of the density fluctuation from the linearized RNS-I can be obtained by solving the following matrix equation \cite{Marques1993}: 
\begin{equation}
\label{eqn_SRBS}
\eqalign{\sf{B}(\omega, {\rm Kn})\, \psi = \sf{D}_s,}
\end{equation}
where
\begin{equation}
\fl{\sf{B} =
\left[\begin{array}{ccc}
-i \omega + 4 \pi^2 \kam^* \rm{Kn} & 2\,\pi i  & 0 \\
2\,\pi i - 8\,\pi^3\,i \left( \frac{4}{3} \kam^* - \kam^{*^2} \right) \rm{Kn}   & - i \omega + 4\,\pi^2 \left(\frac{4}{3} - \kam^* \right) \rm{Kn} & 2\,\pi i  \\
\frac{8}{3}\pi^2 \kam^* \rm{Kn} & \frac{4}{3} \pi i  & -i \omega + \frac{8}{3} \pi^2  \kappa^* \rm{Kn}
\end{array} \right] },
\end{equation}

\begin{equation}
\psi =
\left[\begin{array}{c}
\rho^* \\
U_v^* \\
T^*
\end{array} \right] \quad  \textnormal{and} \quad 
\sf{D}_s =
\left[\begin{array}{c}
1 \\
0 \\
0
\end{array} \right].
\end{equation}
%
%
%
%
%
%

In the coefficient matrix $\sf{B}$, $\omega$ is the angular frequency which is normalized by $\sqrt{\bR\,T_o}/L$. Further, the angular frequency $\omega$ is related to the frequency shift $(f_s)$, which is normalized by a characteristic frequency $\sqrt{2\,\bR\,T_o}/L$. Variables $\rho^*, U_v^*$ and $T^*$ are the spectra of the perturbed density, velocity, and temperature, respectively. The source column matrix $\sf{D}_s$ is due to the initial density perturbation for spontaneous RBS.
For the case of coherent RBS, one has to apply the Fourier transform in both spatial and temporal variables. Then, the coherent RBS spectrum of the density fluctuation can be obtained by solving the following matrix equation:
\begin{equation}
\label{eqn_CRBS}
\eqalign{\sf{B}(\omega, {\rm Kn}) \, \psi = \sf{D}_c \quad \textnormal{with} \quad \sf{D}_c =
\left[\begin{array}{c}
0 \\
1 \\
0
\end{array} \right]},
\end{equation}
where the source column matrix $\sf{D}_c$ is due to the presence of externally induced acceleration for coherent RBS.

%
%
%
%
%
%

By solving the in-homogeneous matrix equations \eref{eqn_SRBS} and \eref{eqn_CRBS} for the spectrum of the density fluctuations $\rho^*$, we obtain the spontaneous RBS and coherent RBS spectra, respectively. Further, quantities $\Re[\rho^*]$ and $|\rho^*|^2 $ describes the spectra for the spontaneous and coherent RBS, respectively. The spontaneous spectra using newly derived hydrodynamic models, namely, RNS-I, RNS-II, RNS-III along with the classical Navier-Stokes predictions are given by:

\begin{equation}
\label{SRBS_RNS}
\eqalign{ \mcs_s^{RNS-I} = \Re[\rho^*] = \Re\left[ \frac{\mathcal{N}^{I}}{\mathcal{D}}\right], \quad & \mcs_s^{RNS-II} = \Re\left[ \frac{\mathcal{N}^{II}}{\mathcal{D}}\right], \\
\mcs_s^{RNS-III} = \Re\left[ \frac{\mathcal{N}^{III}}{\mathcal{D}}\right], \quad & \mcs_s^{NS} = \Re\left[ \frac{\mathcal{N}^{NS}}{\mathcal{D}}\right],}
\end{equation}
where
\begin{eqnarray*}
\fl && \mathcal{N}^{I} = 128\, i\,\kappa^*\,\left(1-\frac{3}{4}\,\kam^*\right) \pi^4\, {\rm{Kn}^2}\, +\, 24\,\left(2+\kappa^*- \frac{3}{2}\,\kam^*\right) \pi^2\, {\rm{Kn}}\, \omega\, - 9 \,i \,\omega^2 \,+\, 24\, i\, \pi^2, \\
\fl && \mathcal{N}^{II} = 128\, i\,\kappa^*\, \pi^4\, {\rm{Kn}^2}\, +\, 24\,(2+\kappa^*) \pi^2\, {\rm{Kn}}\, \omega\, - 9 \,i \,\omega^2 \,+\, 24\, i\, \pi^2 = \mathcal{N}^{NS},\\
\fl && \mathcal{N}^{III} = 128\, i\,\kappa^*\,\left(1-\frac{3}{4}\,\kappa_p^*\right) \pi^4\, {\rm{Kn}^2}\, +\, 24\,\left(2+\kappa^*- \frac{3}{2}\,\kappa_p^*\right) \pi^2\, {\rm{Kn}}\, \omega\, - 9 \,i \,\omega^2 \,+\, 24\, i\, \pi^2,\\
\fl && \mathcal{D} = 128\,\kappa^* \,\pi^4 \,{\rm{Kn}^2} \,\omega \,-\, 24\, i\,(2+\kappa^*) \pi^2\, {\rm{Kn}}\,\omega^2 \,+\, 96\, i\,\kappa^*\, \pi^4\, {\rm{Kn}} \,-\, 9\, \omega^3\, +\, 60\, \pi^2\, \omega.
\end{eqnarray*}

Expression for the coherent RBS spectrum is the same for all models discussed and is given by:
\begin{eqnarray}
&& \mcs_c = \left| \frac{6 \pi  \left(8\,\kappa^*\,\pi^2\,\rm{Kn} \,-\, 3\, i\, \omega \right)}{\mathcal{D}} \right|^2.
\end{eqnarray}

\subsection{Results and interpretation}
\label{4.2}

An RBS spectrum typically consists of a central Rayleigh peak near $f_s = 0$ and two Brillouin side peaks at an equidistance from the central Rayleigh peak. These Brillouin side peaks are located at $f_s = \sqrt{\gamma / 2}$, where $\gamma$ is the ratio of heat capacity. In the typical spectra of the spontaneous RBS, one can identify the contributions from the central Rayleigh peak and the Brillouin side peaks. They are clearly separated from each other when the gas flow is in the hydrodynamic regime ($\rm{Kn} \leq 0.001$). When $\rm{Kn}$ lies in the regime $0.001 \leq \rm{Kn} \leq 0.1$ contributions from the central Rayleigh part and the Brillouin side part become mixed and then the widths of both parts broaden due to the increased dissipation in sound propagation. Further increase of $\rm{Kn}$ results in vanishing contribution of the Brillouin part and the whole spectrum becomes nearly Gaussian \cite{Muller1998,Wu2014,Wu2015,Wu2018}. In the typical spectra of the coherent RBS, one can notice the presence of Brillouin peaks only when the gas flow is in the hydrodynamic regime ($\rm{Kn} \leq 0.001$). As $\rm{Kn}$ increases further, both peaks (the central Rayleigh and the two Brillouin side peaks) are present and the relative intensity of these peaks becomes large and later on the whole spectrum becomes Gaussian \cite{Wu2014,Wu2015,Wu2018}.

\begin{figure}
\centering
\includegraphics[height=6.5cm]{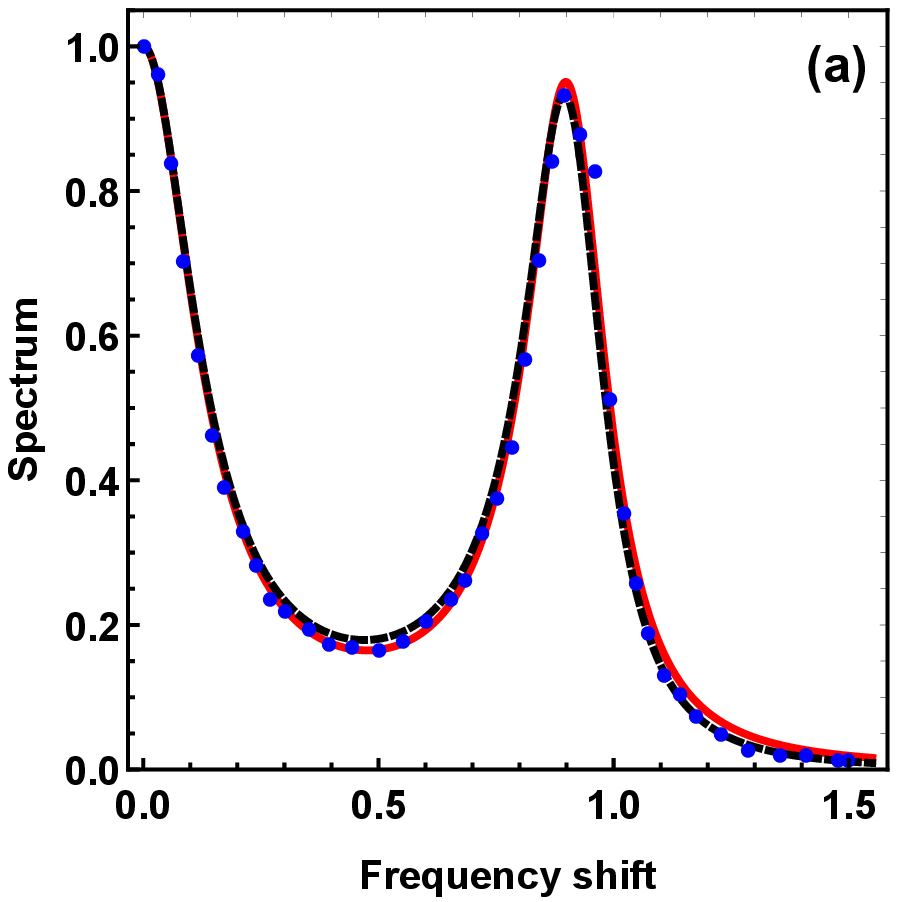}\quad
\includegraphics[height=6.5cm]{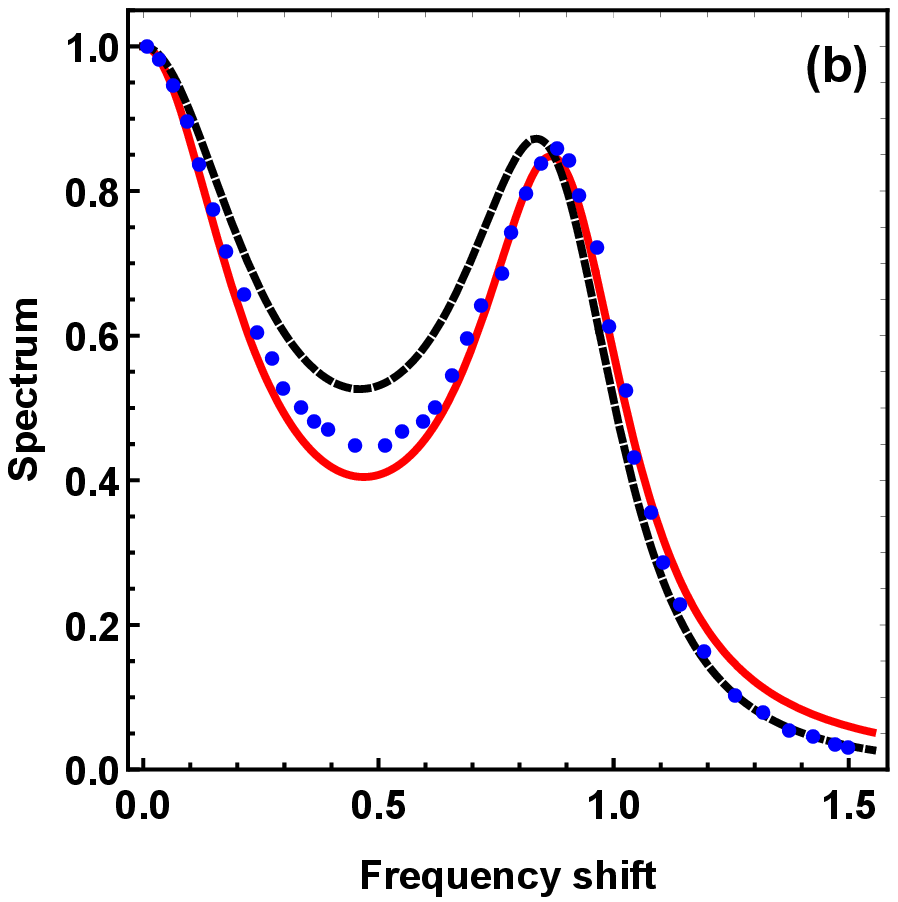}\\
\includegraphics[height=6.5cm]{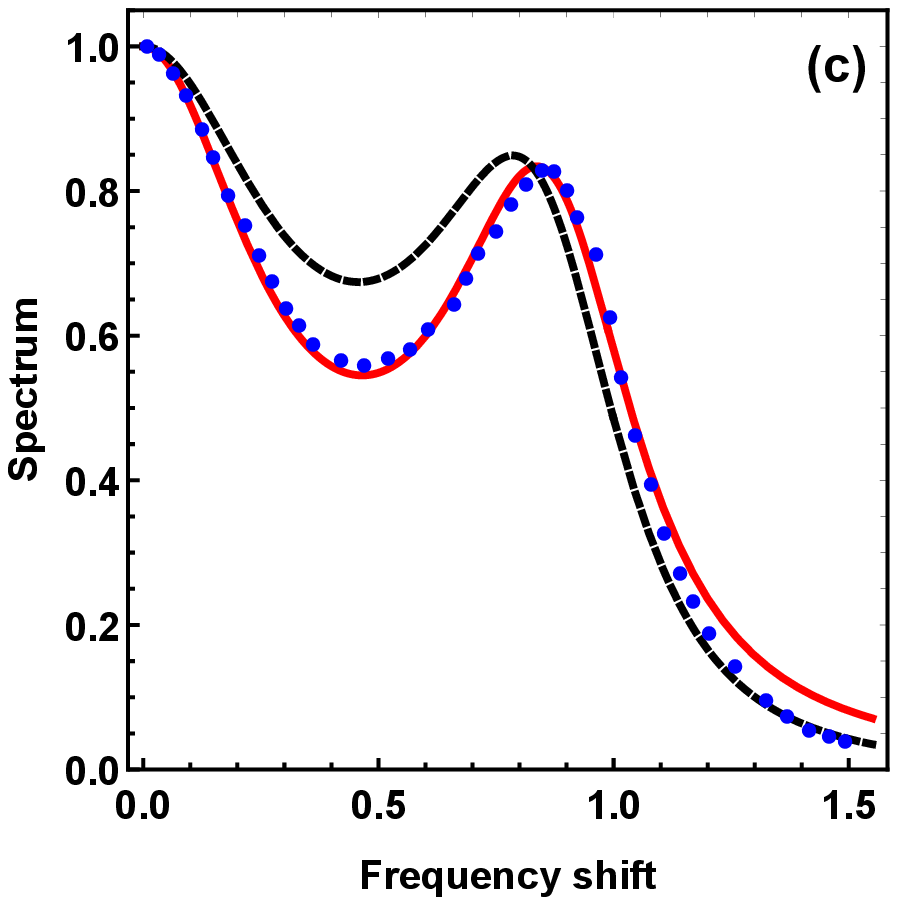}\quad
\includegraphics[height=6.5cm]{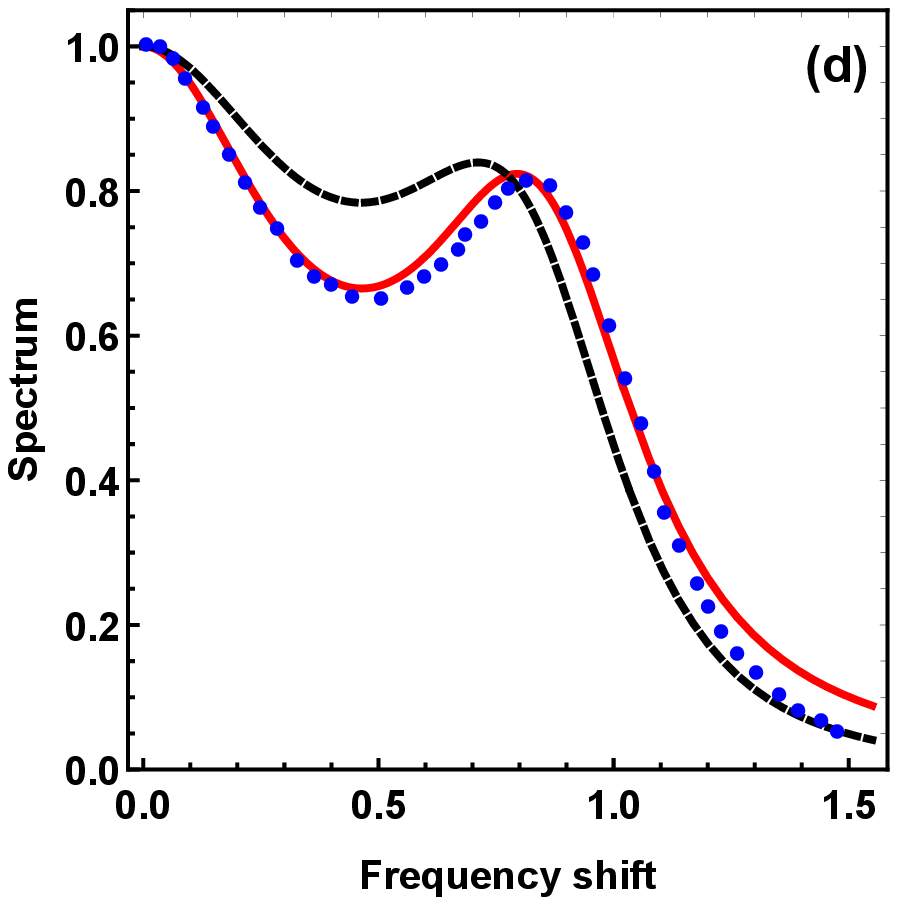}\\
\includegraphics[height=6.5cm]{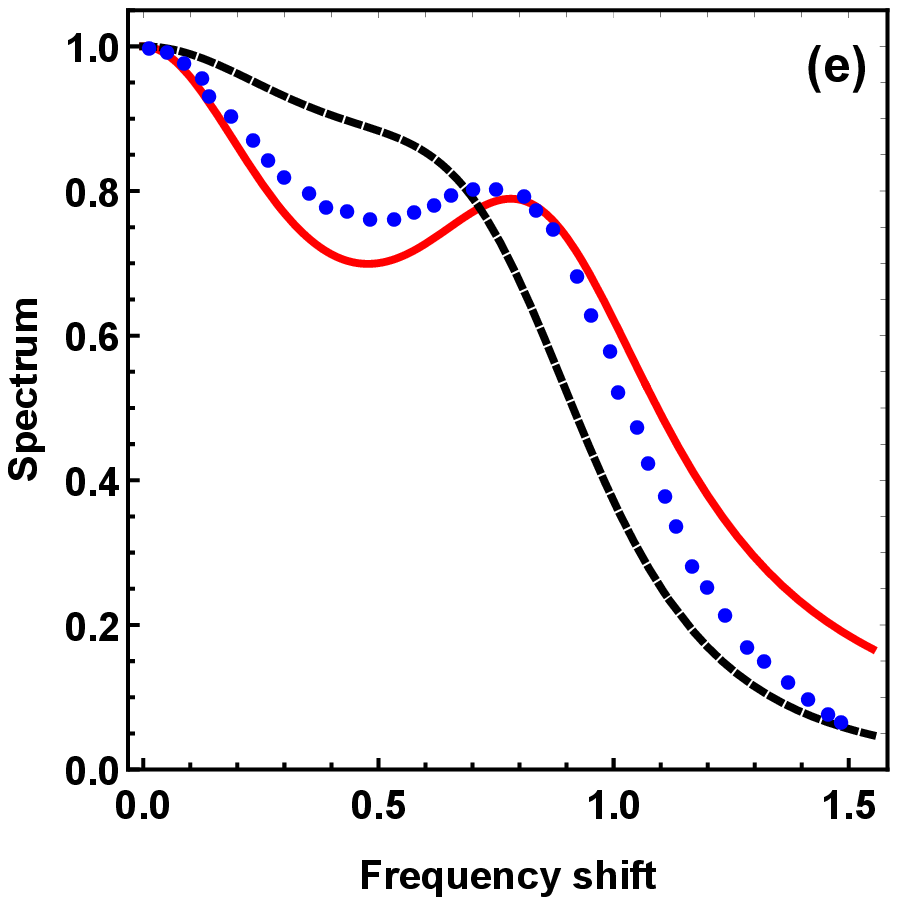}\quad
\includegraphics[height=6.5cm]{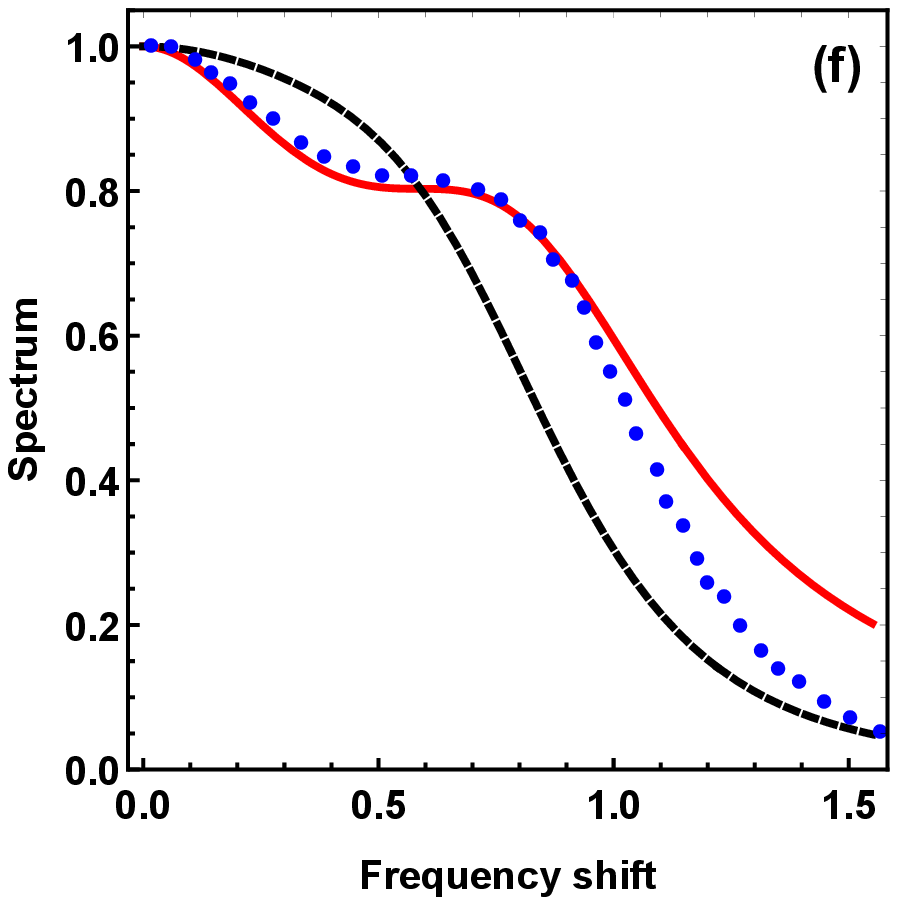}
\caption{Spectra of the spontaneous RBS, when (a) $\rm{Kn} = 0.02, \ka^* = 15/4, \kam^* = 2/5$, (b) $\rm{Kn} = 0.04, \ka^* = 13/4, \kam^* = 1/2$, (c) $\rm{Kn} = 0.05, \ka^* = 13/4, \kam^* = 1/2$, (d) $\rm{Kn} = 0.06, \ka^* = 13/4, \kam^* = 1/2$, (e)  $\rm{Kn} = 0.08, \ka^* = 11/4, \kam^* = 0.85$, and (f)  $\rm{Kn} = 0.1, \ka^* = 11/4, \kam^* = 0.85$. Black dotted line, red solid line and blue filled circles represents the results from the Classical NS, re-casted NS and LBE for Maxwellian gases from Wu~\cite{Wu2018}.}
\label{fig:6}
\end{figure}


\begin{figure}
\centering
\includegraphics[height=6.5cm]{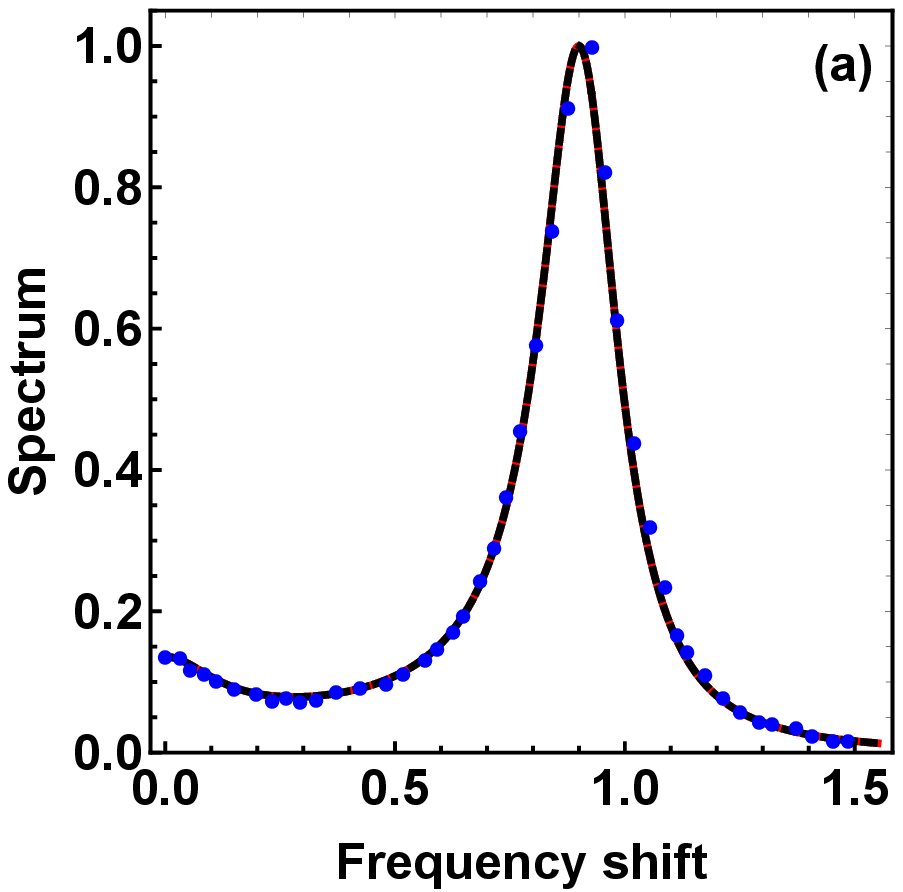}\quad
\includegraphics[height=6.5cm]{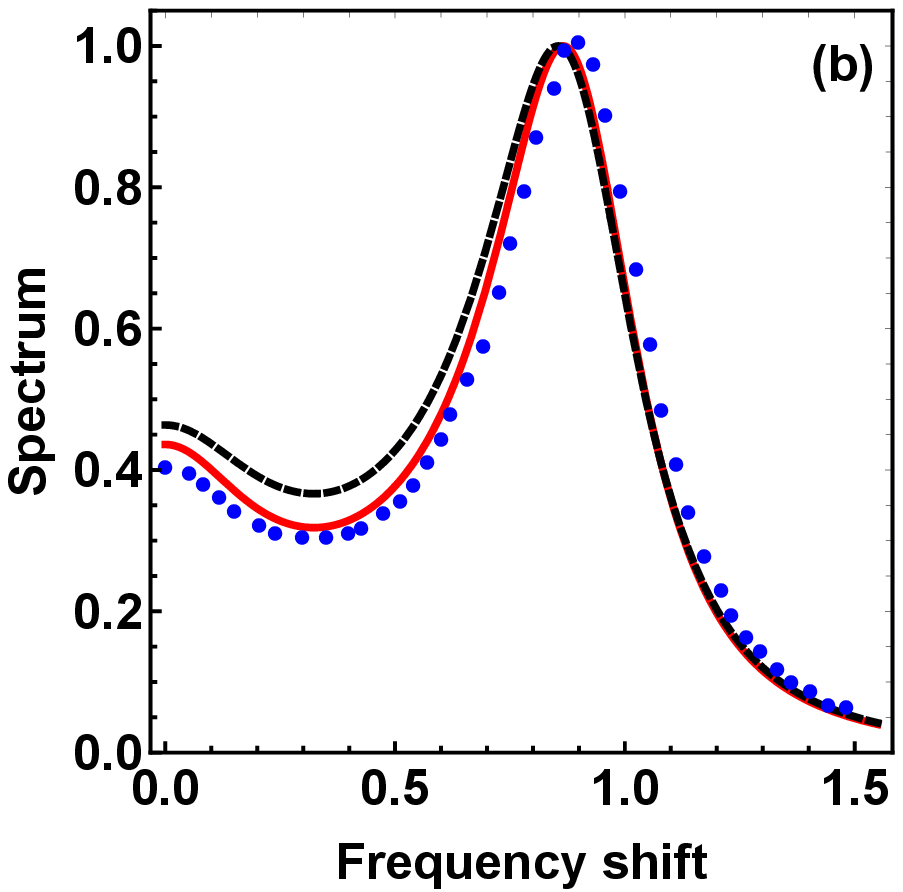}\\
\includegraphics[height=6.5cm]{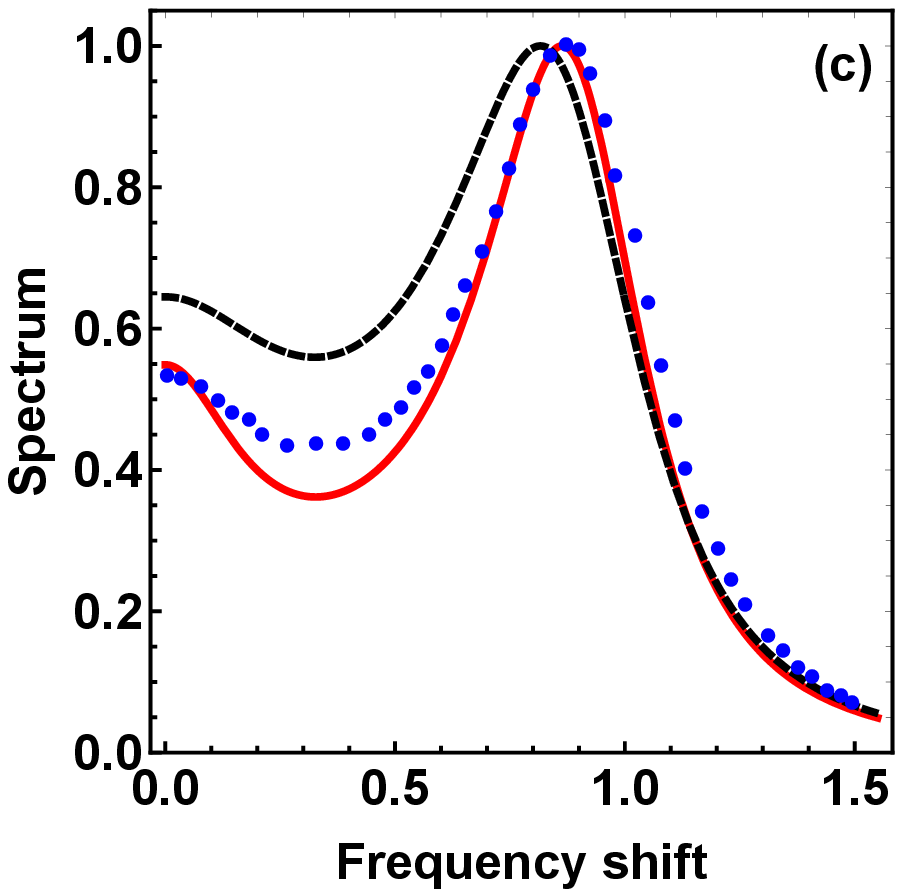}\quad
\includegraphics[height=6.5cm]{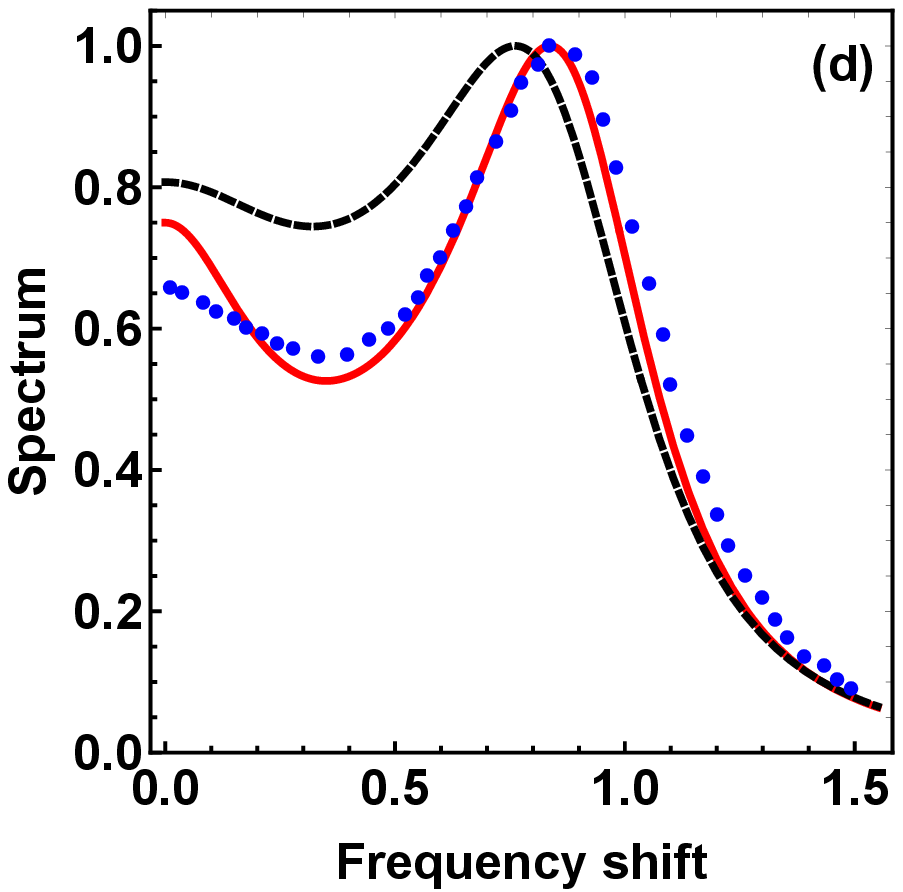}\\
\includegraphics[height=6.5cm]{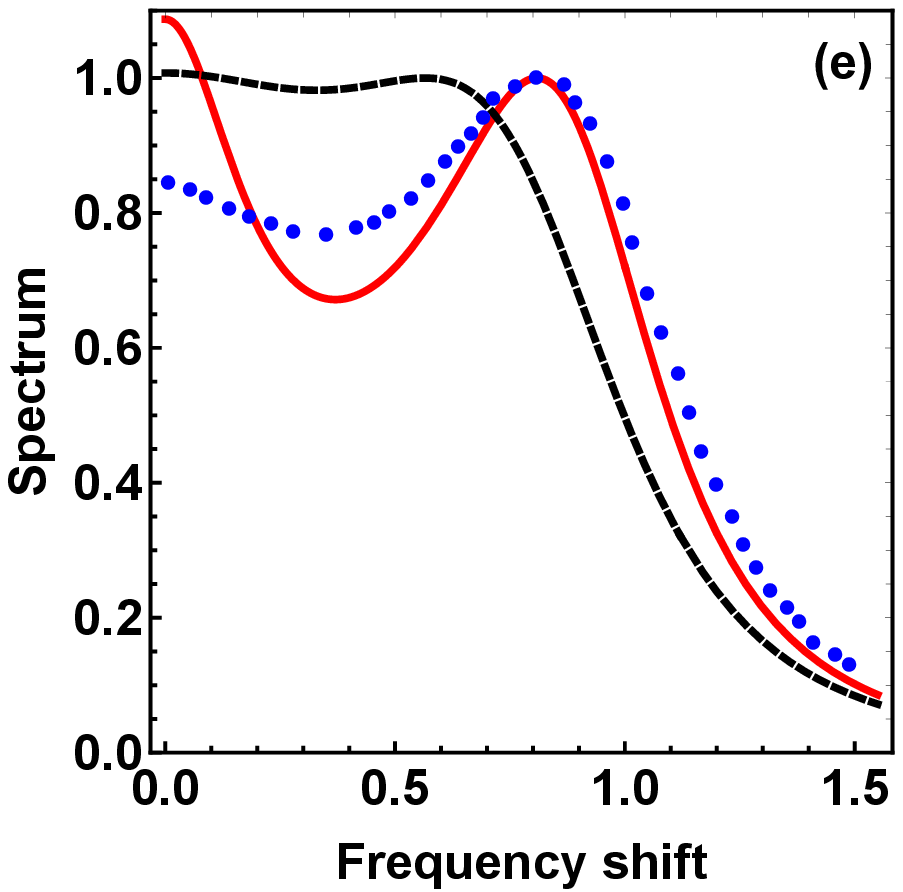}\quad
\includegraphics[height=6.5cm]{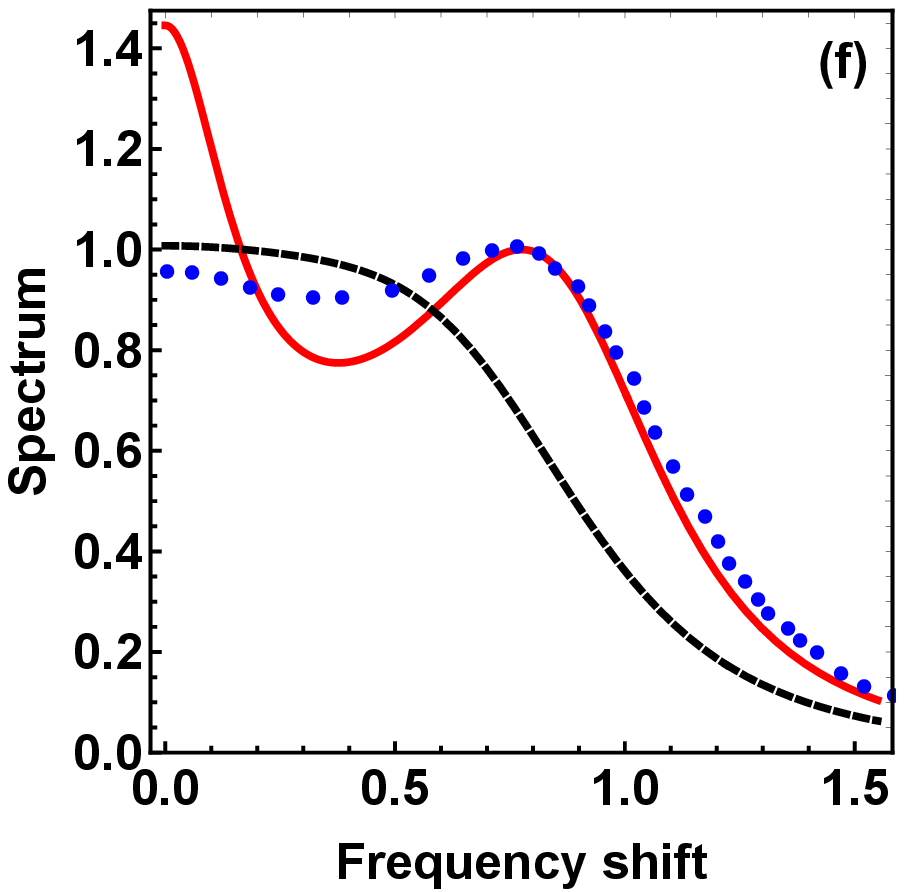}
\caption{Spectra of the coherent RBS when (a) $\rm{Kn} = 0.02, \ka^* = 15/4$, (b) $\rm{Kn} = 0.04, \ka^* = 13/4$, (c) $\rm{Kn} = 0.05, \ka^* = 9/4$, (d) $\rm{Kn} = 0.06, \ka^* = 9/4$, (e) $\rm{Kn} = 0.08, \ka^* = 3/2$, and (f) $\rm{Kn} = 0.1, \ka^* = 1$. Black dotted line, red solid line and blue filled circles represents the results from the Classical NS, re-casted NS and LBE for Maxwellian gases from Wu~\cite{Wu2018}.}
\label{fig:7}
\end{figure}


From the spontaneous spectrum expression \eref{SRBS_RNS}, which is predicted by our re-casted Navier-Stokes, we observe that the spontaneous RBS spectra depends explicitly on the Knudsen number, the new molecular diffusivity coefficients, and the thermal conductivity coefficient.
The coherent RBS spectra expression is the same for all models and appears to depend only on the Knudsen number, $\rm{Kn}$, and the thermal conductivity coefficient $\kappa^*$. As transport coefficients associated with volume/mass diffusion theory are different from those in the classical theory, we here compare the spectral lines results from re-casted Navier-Stokes models by considering transport coefficients which give best match with the LBE spectral lines. A coherence is found as those best coefficients appear to depend on the Knudsen number and listed later in Tables \eref{tab:SRBS} and \eref{tab:CRBS}. Classical Navier-Stokes coefficients are kept to their exact values, i.e., $\kappa^*=15/4$. RBS spectrum is symmetric about the position of the central Rayleigh peak, $f_s = 0$. So we only plot and compare the half of the spectrum  corresponding to the positive frequency shift $f_s$. Furthermore, in all figures, the RBS spectrum has been normalized by the maximum value.

First, we compare the spontaneous RBS spectra solutions obtained from the re-casted Navier-Stokes with that of the classical Navier-Stokes solutions and also with the results calculated based on the Linearized Boltzmann equation (LBE) for Maxwellian gases and taken from Wu~\cite{Wu2018}. Figure \ref{fig:6} illustrates the spontaneous RBS spectra obtained from the re-casted Navier-Stokes-I model along with the spectra obtained from the classical Navier-Stokes and the LBE. In figure \ref{fig:6}, panel (a), (b), (c), (d), (e) and (f) we show the spontaneous spectra results for $\rm{Kn} = 0.02, 0.04, 0.05, 0.06, 0.08$ and $0.1$, respectively, in which the black dotted line, the red solid line and the blue filled circles correspond to the solution by the classical NS, the re-casted NS and the LBE solutions \cite{Wu2018}, respectively. When $\rm{Kn} = 0.02$, we found that the re-casted Navier-Stokes spectrum have an excellent agreement with the LBE spectrum line for the choice of transport coefficients $\ka^*$ and $\kam^*$ to be $15/4$ and $2/5$, which can be seen from figure \ref{fig:6} (a). At this Knudsen number, classical Navier-Stokes also predicts the actual spectrum line.

\begin{table}
\begin{center}
\caption{\label{tab:SRBS}Spontaneous RBS spectra: Knudsen number vs values of different transport coefficients for which the re-casted NS models have a best fit with the LBE Spectral lines.}
\item[]\begin{tabular}{@{}lllll}
\br
 $\rm{Kn}$  & $\ka^*$   &   $\kam^*$ & $\kat^*$ & $\kap^*$ \\
\mr
       0.02  & 15/4 & 2/5  & -- & 2/5  \\
       0.04  & 13/4 & 1/2  & -- & 1/2 \\
       0.05  & 13/4 & 1/2  & -- & 1/2 \\
       0.06  & 13/4 & 1/2  & -- & 1/2 \\
       0.08  & 11/4 & 0.85 & -- & 0.85 \\
       0.1   & 11/4 & 0.85 & -- & 0.85 \\
\br
\end{tabular}
\end{center}
\end{table}
Typical profiles of the spontaneous RBS spectra for $\rm{Kn} = 0.04, 0.05$ and $0.06$ are shown in figure \ref{fig:6} (b), figure \ref{fig:6} (c) and figure \ref{fig:6} (d), respectively. From these spectral solutions, we observe that the re-casted Navier-Stokes solutions achieved the best  agreement with that of LBE solutions with $\ka^* = 13/4$ and $\kam^* = 1/2$, while classical Navier-Stokes solutions deviate significantly from the LBE solutions. Moreover, it is noteworthy to mention that classical Navier-Stokes predicts the higher spectrum near the Brillouin side peaks whereas our re-casted Navier-Stokes (RNS-I and RNS-III) are successful in predicting the actual spectrum like LBE solutions. Figures \ref{fig:6} (e) and (f) show the profiles of the spontaneous RBS for $\rm{Kn} = 0.08$ and $\rm{Kn} = 0.1$, respectively. Even at this higher Knudsen numbers, the predictions of re-casted Navier-Stokes with the choice of $\ka^* = 11/4$ and $\kam^* = 0.85$ have showed better agreement with that of LBE solutions, while the classical Navier-Stokes fails to predict the spectrum. Overall, one can conclude from Figures \ref{fig:6} that the re-casted Navier-Stokes models, namely, RNS-I and RNS-III performed better in predicting the spontaneous spectrum as compared with LBE spectrum up to $\rm{Kn} = 0.1$ with coefficients as listed in Table \ref{tab:SRBS}. Figure \ref{fig:7} shows the typical shapes of the coherent RBS spectra obtained from the re-casted NS and classical NS models. In all panels, the black dotted and red solid lines represent the solutions from classical Navier-Stokes and re-casted NS and blue filled circles represent the LBE solutions from \cite{Wu2018}. In figure \ref{fig:7}, panels (a), (b), (c), (d), (e) and (f) show the spectrum solutions for $\rm{Kn} = 0.02, 0.04, 0.05, 0.06, 0.08$ and $0.1$, respectively, with the corresponding thermal conductivity coefficient $\ka^*$ listed in Table \ref{tab:CRBS}. Like in the case of the spontaneous spectra, the classical Navier-Stokes model performs well up to $\rm{Kn} = 0.02$, as evident from figure \ref{fig:7} (a). It is customary to conclude from figure \ref{fig:7} (a) that at $\rm{Kn} = 0.02$, both classical NS and re-casted NS models shows a perfect agreement with the LBE solutions. When $\rm{Kn} = 0.04, 0.05$ and $0.06$, the coherent spectra solutions from re-casted NS models with respectively associated $\ka^* = 13/4, 9/4$ and $9/4$, yield better agreement with the LBE solutions, than the classical NS model, which is evident from figure \ref{fig:7} (b),  figure \ref{fig:7} (c), and figure \ref{fig:7} (d), respectively. At these Knudsen numbers, one can observe from the classical NS solution that the position of the Brillouin peak shifted towards the left as compared to the LBE and re-casted NS solutions but qualitatively predicts the spectral line shape as that of the LBE solution. Observing the coherent RBS spectral solutions at $\rm{Kn} = 0.08$ and $0.1$ presented in figure \ref{fig:7} (e) and figure \ref{fig:7} (f), respectively, one can conclude that classical NS model will not be able to produce the actual shape of the spectral lines but re-casted NS model are still successful in predicting the actual shape of the coherent RBS spectrum. Re-casted NS models, however, predict higher spectrum at the central Rayleigh part than that of LBE predictions and are able to predict the Brillouin part contributions accurately.

Based on the spontaneous and coherent spectral lines presented in figure \ref{fig:6} and figure \ref{fig:7}, one can conclude as follows: for the case of spontaneous RBS, RNS-I and RNS-III models perform well up to $\rm{Kn} \approx 0.1$. Classical NS model seems to perform well up to $\rm{Kn} \approx 0.02$ only (see figure \ref{fig:6}). For the case of coherent RBS, like in the case of the spontaneous RBS, the accuracy of classical NS model is limited to $\rm{Kn} \approx 0.02$ only. All re-casted NS models perform well up to $\rm{Kn} \approx 0.06$ and also shows better agreement with LBE solutions up to $\rm{Kn} \approx 0.1$. Overall, one can say that the re-casted Navier-Stokes equations are successful in predicting the shape of the RBS spectral lines up to $\rm{Kn} \approx 0.1$.


\begin{table}
\begin{center}
\caption{\label{tab:CRBS}Coherent RBS spectra: Knudsen number vs values of thermal conductivity coefficient for which the re-casted NS models have a best fit with the LBE Spectral lines.}
\item[]\begin{tabular}{@{}lllllll}
\br
  $\rm{Kn}$  & 0.02  & 0.04  & 0.05 & 0.06 & 0.08 & 0.1 \\[3pt]
\mr
  $\ka^*$     & 15/4  & 13/4  & 9/4  & 9/4  & 3/2  & 1    \\
\br
\end{tabular}
\end{center}
\end{table}

\section{Discussion}
\label{5}

In order to improve accuracy of the original fluid flow equations, extra terms are usually constructed to modify constitutive equations for the shear stress tensor and/or the heat flux vector only. Chapman-Enskog expansion to obtain solutions to the Boltzmann equation is typically the method used to obtain high order equations for rarefied gas flows. However, constitutive equations obtained in this manner are well-known to lead to equations that violate mechanical properties or the second law. The methodology introduced in this article is a systematic method involving the three conservation equations combined. Our three new constructed continuum flow models can be reverted by the following trivial change of variables:
\begin{equation}
\label{eqn_MDvel}
U_v = U + \kam \nabla \ln \rho \ ,
\end{equation}
\begin{equation}
\label{eqn_TDvel}
U_T = U + \kat \nabla \ln T \ ,
\end{equation}
and
\begin{equation}
\label{eqn_PDvel}
U_p = U + \kap \nabla \ln p.
\end{equation}
Substituting equation \eref{eqn_MDvel}, \eref{eqn_TDvel} and \eref{eqn_PDvel} into the continuum flow system \eref{eqn_RNSmass} - \eref{eqn_RNSenergy}, \eref{eqn_TRNSmass} - \eref{eqn_TRNSenergy} and \eref{eqn_PRNSmass} - \eref{eqn_PRNSenergy}, respectively, by reversing the procedure as described in the Appendices, is expected to lead back to the system \eref{eqn_mass} - \eref{eqn_energy} they originated from. This original system satisfies all known mechanical properties \cite{Woods1993}. It may therefore be concluded, at first, that all our three re-casted models are fully thermo-mechanically consistent equations via the original. The new systems exhibit new physics (e.g., Korteweg shear stress and Sone's Ghost effect stress) not seen in the original equations. Subsequently, while the original Navier-Stokes and its transformed version may be  mathematically convertible from one to the other, they do display different physics. Comparing solutions of the transformed equations in terms of to experiments do not systematically equals comparing solutions of the original to experimental data. This is demonstrated here by our comparisons with the Rayleigh-Brillouin light scattering experiments.
It serves as a direct theoretical support to Brenner's observation of the experimental difference between a dye- or photochromic experiments (measuring a fluid's mass velocity) and a tracer velocity \cite{HBrenner2009}, which led him to initiate the first Bi-velocity hydrodynamics theory. As it was already observed that continuum flow equations written with explicit diffusive component in the continuity violates mechanical properties \cite{DadzieReese2012}, we conclude that the form of the equations that should be used to analyse their mechanical properties is their equivalent form written in terms of the mass velocity. In this particular case, the original system \eref{eqn_mass} - \eref{eqn_energy} which already satisfies those properties. In other words, differentiating properly between the different type of velocities is also primordial to fully analyse the flow equation mechanical properties as listed in {\"O}ttinger \etal.\cite{Ottinger2009}. An advantage of the present strategy to construct new continuum flow models may therefore be maintaining the conservation of those mechanical properties via the original Navier-Stokes equations.

The transformation technique presented to obtain the new models made use of the three basic thermodynamics variables, namely, density, temperature and pressure. The methodology can be extended by combining these variables or form more complex change of variables to obtain more complex and systematically thermo-mechanically consistent continuum flow models. For example, using the following change of variable:
\begin{equation}
 U = U_{\tau} - \ka_{\tau} \, \nabla \times U_{\tau},
\end{equation}
will incorporate fluid vorticity contributions to the shear stress tensor.
Then, it is trivial that the accompanying mass diffusion in the continuity equation will be driven by fluid vorticity.
Burnett and super-Burnett equations for rarefied gases involve complex coupling terms in shear stress and heat flux, for example, coupling between velocity gradient and temperature gradient etc., \cite{Woods1993}. To produce a re-casted Navier-Stokes that introduces high order terms of the type of Burnett or super-Burnett terms, the following types of velocity transformations may be constructed:
\begin{eqnarray}
 U = U_{\pi T} - \frac{\kat}{T} \nabla T \cdot \bPi_{\pi T}; \quad
 U = U_{\pi p} - \frac{\kap}{p} \nabla p \cdot \bPi_{\pi p},
\end{eqnarray}
with
\begin{eqnarray}
 \bPi_{\pi T} =  -2\, \mu\, \mathring{\overline{\nabla U_{\pi T}}}   ;\quad
 \bPi_{\pi p} =  -2\, \mu\, \mathring{\overline{\nabla U_{\pi p}}}.
\end{eqnarray}
Finally, although the original Navier-Stokes equations is used to demonstrate the new strategy, the methodology itself is applicable to other continuum equations with original strong mechanical properties.

\section{Conclusion}
\label{6}

We have introduced a new framework based on a transformation of the velocity vector field within the standard Navier-Stokes equations to obtain new continuum flow equations of volume/mass diffusion types. The new continuum flow equations termed the re-casted Navier-Stokes equations are systematically thermo-mechanically consistent via the original. All three re-casted forms of the classical Navier-Stokes are fully parabolic systems unlike the classical one, in which the absence of diffusion term in the continuity equation is responsible for not being fully parabolic.
The new equations also display various important physics that the original lacks.  Based on the plane wave analysis, we confirm that the dispersion relation for all re-casted NS equations is the same and it coincides with that of the classical Navier-Stokes dispersion relation; hence all re-casted NS models are shown to be both temporally and spatially stable. Our analysis of the Rayleigh-Brillouin scattering experimental data demonstrated that the re-casted Navier-Stokes equations are capable of describing the RBS spectrum shapes better than the untransformed equations. This comes in support for potential existence of various meaningful different fluid flow velocities. As future works we will apply these new re-casted Navier-Stokes models to test other configurations where the original fails such as in the description of shock wave structures. The methodology will also be deployed to obtain new flow equations to other problems such as particle-laden flows.

\section{Acknowledgments}

This research is supported by the UK's Engineering and Physical Sciences Research Council (EPSRC) under grant nos. EP/N016602/1, EP/R007438/1, EP/R008027/1, and EP/N034066/1. JMR acknowledges the support of the Royal Academy of Engineering under the Chair in Emerging Technologies scheme. SKD acknowledges the support of The Leverhulme Trust under grant Ref. RPG-2018-174.


\appendix
\section{Useful definitions and some vector identities\label{Appendix_A}}
\begin{enumerate}
 \item The Hessian operator/matrix is denoted by $\widetilde{\bD}$ and is defined by
 \begin{equation}
 \label{eqn_A1}
 \widetilde{\bD} f = \frac{\partial }{\partial X_i} \frac{\partial }{\partial X_j} f = \frac{\partial^2 f}{\partial X_i \partial X_j}.
 \end{equation}
 For continuous scalar field $f$, the order of the differentiation does not matter.

  \item The tensor (dyadic) product of two vectors $\mathbf{a}$ and $\mathbf{b}$ is denoted by $\mathbf{a} \otimes \mathbf{b}$ and is defined as

 \begin{equation}
 \label{eqn_A2}
 \mathbf{a} \otimes \mathbf{b} = \mathbf{a} \mathbf{b}^T = \mathbf{a}_i \mathbf{b}_j =
 \left[\begin{array}{ccc}
    a_1b_1 & a_1b_2 & a_1b_3\\
    a_2b_1 & a_2b_2 & a_2b_3\\
    a_3b_1 & a_3b_2 & a_3b_3
 \end{array} \right]
 \end{equation}

 \item The divergence of dyadic product of two vectors $\mathbf{a}$ and $\mathbf{b}$ is given by
  \begin{equation}
 \label{eqn_A3}
 \nabla \cdot( \mathbf{a} \otimes \mathbf{b}) =  \mathbf{b} (\nabla \cdot \mathbf{a}) + (\mathbf{a} \cdot \nabla) \mathbf{b}
 \end{equation}

 \item For any scalar filed $f$ and vector field $\mbfF$
 \begin{equation}
 \label{eqn_A4}
 \nabla \cdot(f \mbfF) =  f \nabla \cdot \mbfF + \mbfF \cdot \nabla f
 \end{equation}

 \item Useful Identities:

  \begin{equation}
 \label{eqn_A5}
 - 2 \mu \mathring{\overline{\nabla U_v}} = -\,2 \,\mu \,\mathbf{D}\left(U_v\right)\,-\,\lambda\, \left( \nabla \cdot U_v\right)\, \bI,     \end{equation}

 \begin{equation}
 \label{eqn_A6}
 \widetilde{\bD} \ln \rho = - \frac{1}{\rho^2}\, \nabla \rho \otimes \nabla \rho \, +\, \frac{1}{\rho}\,\widetilde{\bD} \rho,
 \end{equation}

 \begin{equation}
 \label{eqn_A7}
 \Delta \ln \rho =  \frac{1}{\rho}\,\Delta \rho \, -\, \frac{|\nabla \rho|^2}{\rho^2}.
 \end{equation}

 \item $U^2$ in terms of $U_v$
 \begin{eqnarray}
 \label{eqn_A8}
 \fl U^2 &&=  U_v^2 \,-\, 2\, \kam \,(U_v \cdot \nabla \ln \rho)\, +\, \kam^2 \,\left(\nabla \ln \rho \cdot \nabla \ln \rho \right) \quad \textnormal{where} \, \nabla = \frac{\partial }{\partial X_i}
 \end{eqnarray}

 \item We remark that
 \begin{eqnarray}
 \label{eqn_A9}
 \nabla  (U_v \cdot \nabla \rho) - \frac{\nabla \rho}{\rho} (U_v \cdot \nabla \rho) - \rho (U_v \cdot \nabla) \frac{\nabla \rho}{\rho} = \widetilde{\nabla U_v} \cdot \nabla \rho
 \end{eqnarray}
 Proof: Consider the quantity
 \begin{equation}
 \fl \eqalign{& \nabla  (U_v \cdot \nabla \rho) - \frac{\nabla \rho}{\rho} (U_v \cdot \nabla \rho) - \rho (U_v \cdot \nabla) \frac{\nabla \rho}{\rho} \\
 & = \frac{\partial }{\partial X_i} \left(U_{v_j} \frac{\partial \rho}{\partial X_j} \right) - \frac{1}{\rho} \frac{\partial \rho}{\partial X_i} \left( U_{v_j} \frac{\partial \rho}{\partial X_j} \right) - \rho \left( U_{v_j} \frac{\partial }{\partial X_j} \right) \frac{1}{\rho} \frac{\partial \rho}{\partial X_i}\\
 & =\frac{\partial U_{v_j}}{\partial X_i}  \frac{\partial \rho}{\partial X_j} + U_{v_j} \widetilde{\bD} \rho - \frac{1}{\rho} U_{v_j} \frac{\partial \rho}{\partial X_i} \frac{\partial \rho}{\partial X_j} - \rho U_{v_j} \left(\frac{-1}{\rho^2}\right) \frac{\partial \rho}{\partial X_j} \frac{\partial \rho}{\partial X_i} - U_{v_j} \widetilde{\bD} \rho\\
 & = \frac{\partial U_{v_j}}{\partial X_i}  \frac{\partial \rho}{\partial X_j} = \widetilde{\nabla U_v} \cdot \nabla \rho}
 \end{equation}

\end{enumerate}

\section{Re-casting the momentum balance equation}
\label{Appendix_B}
The conservative form of the classical momentum balance equation is given by
\begin{eqnarray}
\label{eqn_B1}
\frac{\partial  \rho \, U }{\partial t}  \,+  \, \nabla \cdot [\rho \,U \otimes U]\,+ \, \nabla \cdot [p \,\pmb{I} \,+\, \bPi^{(NS)}] = 0.
\end{eqnarray}

Here, we present the detailed algebra involved while re-casting the above momentum balance equation \eref{eqn_B1} using the relation given in   \eref{eqn_mvel}. The first term in \eref{eqn_B1} can be transformed as:
\begin{equation}
\fl \eqalign{ \frac{\partial  \rho \, U }{\partial t} &= \frac{\partial  }{\partial t} \left[  \rho \, U_v - k_m \rho \, \nabla \ln \rho \right] =  \frac{\partial \rho \, U_v }{\partial t}  - \kam \frac{\partial  }{\partial t}  \nabla \rho \\
&= \frac{\partial \rho \, U_v }{\partial t}  - \kam  \nabla  \frac{\partial \rho }{\partial t} \qquad [\because\textbf{As X and t are independent variables}]\\
&= \frac{\partial \rho \, U_v }{\partial t}  - \kam  \nabla  \left[ -\, \nabla \cdot (\rho \, U_v)\, + \,\kam \, \Delta \rho \right] \qquad [\because Equation  \eref{eqn_rmass1}] \\
&= \frac{\partial \rho \, U_v }{\partial t}\,+ \, \kam \, \nabla \left[ \nabla \cdot (\rho \, U_v ) \right]\,-\, \kam^2  \nabla \Delta \rho}
\end{equation}

\begin{equation}
\label{eqn_B2}
\fl \eqalign{\therefore \, \frac{\partial \rho \, U }{\partial t} &= \frac{\partial \rho \, U_v }{\partial t}\,+ \, \kam \, \nabla \left[ \nabla \cdot (\rho \, U_v ) \right]\,-\, \kam^2  \nabla \Delta \rho}
\end{equation}

Similarly, the term  $\nabla \cdot [\rho \,U \otimes U]$ on L.H.S of \eref{eqn_B1} can be transformed as:
\begin{equation}
\fl \eqalign{ \nabla \cdot [\rho \,U \otimes U] \\
= \nabla \cdot \left[ \rho \left( U_v - \kam \nabla \ln \rho \right) \otimes \left( U_v - \kam \nabla \ln \rho \right) \right]\\
= \nabla \cdot \left[ \rho \,U_v \otimes U_v - \rho \kam U_v \otimes \nabla \ln \rho  - \rho \kam  \nabla \ln \rho \otimes U_v + \rho \kam^2 \nabla \ln \rho \otimes \nabla \ln \rho \right]\\
= \nabla \cdot \left[ \rho \,U_v \otimes U_v - \kam U_v \otimes \nabla \rho  - \kam  \nabla \rho \otimes U_v + \frac{\kam^2}{\rho} \nabla \rho \otimes \nabla \rho \right]}
\end{equation}

\begin{equation}
\label{eqn_B3}
\fl \eqalign{ \therefore \nabla \cdot [\rho \,U \otimes U] &= \nabla \cdot \left[ \rho \,U_v \otimes U_v - \kam U_v \otimes \nabla \rho - \kam  \nabla \rho \otimes U_v + \frac{\kam^2}{\rho} \nabla \rho \otimes \nabla \rho \right]}
\end{equation}

The classical shear stress tensor $\bPi^{(NS)}$ is transformed to $\bPi_v$ (classical shear stress in terms of volume velocity) as
\begin{eqnarray*}
\fl \bPi^{(NS)} &=& - 2 \mu \left[ \frac{1}{2} (\nabla U + \widetilde{\nabla U}) - \frac{1}{3} \bI \nabla \cdot U \right] = - 2 \mu \left[ \frac{1}{2} \left( \frac{\partial U_i}{\partial X_j} + \frac{\partial U_j}{\partial X_i}  \right) - \frac{1}{3} \delta_{ij} \frac{\partial U_k}{\partial X_k}\right]\\
\fl &=& - 2 \mu \left[ \frac{1}{2} \left( \frac{\partial U_{v_i}}{\partial X_j} - \kam \widetilde{\bD} \ln \rho + \frac{\partial U_{v_j}}{\partial X_i} - \kam \widetilde{\bD} \ln \rho \right) - \frac{\delta_{ij}}{3}  \left( \frac{\partial U_{v_k}}{\partial X_k} + \kam \Delta \ln \rho \right) \right]\\
\fl &=& - 2 \mu \left[ \frac{1}{2} \left( \frac{\partial U_{v_i}}{\partial X_j} + \frac{\partial U_{v_j}}{\partial X_i} \right) - \frac{1}{3} \delta_{ij} \frac{\partial U_{v_k}}{\partial X_k} \right] \,+ \,2 \,\mu \,\kam \,\widetilde{\bD} \ln \rho - \frac{2\,\mu}{3} \delta_{ij}\, \kam \,\Delta \ln \rho \\
\fl &=& - 2 \mu \mathring{\overline{\nabla U_v}}\,+ \, 2\, \mu \, \kam \widetilde{\bD} \ln \rho - \frac{2\, \mu}{3} \delta_{ij} \,\kam \,\Delta \ln \rho \,=\, \bPi_v.
\end{eqnarray*}

\begin{eqnarray}
\label{eqn_B4}
\therefore \, \bPi^{(NS)} \rightarrow  \bPi_v  &=& - 2 \mu \mathring{\overline{\nabla U_v}}\, + \, 2\, \mu \, \kam \widetilde{\bD} \ln \rho \, + \, \lambda \,\kam \, \Delta \ln \rho \, \bI.
\end{eqnarray}

Using \eref{eqn_B4}, the last term on L.H.S of \eref{eqn_B1} can be transformed into
\begin{equation}
\label{eqn_B5}
\nabla \cdot \left[ p \bI + \bPi_v \right].
\end{equation}

The re-casted form of the momentum balance equation \eref{eqn_B1} is
\begin{eqnarray}
\label{eqn_B6}
\fl &&  \frac{\partial }{\partial t} \left( \rho \, U_v\,- \, \kam\, \nabla \rho \right)\,+\, \nabla \cdot \left[ \rho \,U_v \otimes U_v - \kam U_v \otimes \nabla \rho - \kam  \nabla \rho \otimes U_v + \frac{\kam^2}{\rho} \nabla \rho \otimes \nabla \rho \right] \nonumber \\
\fl && \qquad \qquad \qquad \qquad +\, \nabla \cdot \left[ p \, \bI + \bPi_v\,\right] = 0.
\end{eqnarray}
Let us assume that
\begin{eqnarray}
\label{eqn_B7}
\bPi^{(RNS)}_v &=& \bPi_v\,- \kam U_v \otimes \nabla \rho - \kam  \nabla \rho \otimes U_v + \frac{\kam^2}{\rho} \nabla \rho \otimes \nabla \rho,
\end{eqnarray}
then the final form of the re-casted momentum balance equation gets the following form.
\begin{eqnarray}
\label{eqn_B8}
\fl && \frac{\partial }{\partial t} \left( \rho \, U_v\,- \, \kam\, \nabla \rho \right)\,+\, \nabla \cdot \left[ \rho \,U_v \otimes U_v\right]\,+\, \nabla \cdot \left[  p \, \bI \,+\, \ \bPi^{(RNS)}_v \right] = 0, \nonumber \\
\fl \mbox{or}\\
\fl && \frac{\partial \rho \, U_v }{\partial t}\,+\, \nabla \cdot \left[ \rho \,U_v \otimes U_v\right]+\nabla \cdot \left[ p \,\bI + \bPi^{(RNS)}_v \right]+ \kam \, \nabla \left[ \nabla \cdot (\rho \, U_v ) \right]- \kam^2  \nabla \Delta \rho = 0. \nonumber
\end{eqnarray}
\section{Non-conservative form of re-casted momentum equation}
\label{Appendix_C}
The non-conservative form of the classical momentum balance equation \eref{eqn_B1} is given by:
\begin{equation}
\label{eqn_C1}
\rho \left[ \frac{\partial  U }{\partial t}  +  (U \cdot \nabla) \,U \right]\,+ \,\nabla \cdot [p \pmb{I} + \bPi^{(NS)}] = 0.
\end{equation}
Let us consider the first term in \eref{eqn_C1} which can be transformed as:
\begin{equation}
\fl \eqalign{ \rho \, \frac{\partial  U}{\partial t} &=  \rho \, \frac{\partial U_v}{\partial t}  \,- \,\kam \,\rho\,\nabla  \left( \frac{\partial \ln \rho}{\partial t}  \right),\qquad (\because\mbox{As $X$ and $t$ are independent variables})\\
\fl &=  \rho\,  \frac{\partial U_v }{\partial t} \, + \,\kam \,\frac{\nabla \rho}{\rho} \frac{\partial  \rho}{\partial t} \,-\, \kam \, \nabla  \left(\frac{\partial \rho}{\partial t}\right),\\
\fl &=  \rho  \frac{\partial U_v }{\partial t}  + \kam \frac{\nabla \rho}{\rho} \left[ \kam \Delta \rho - \nabla \cdot (\rho \, U_v )\right] - \kam  \nabla   \left[ \kam \Delta \rho - \nabla \cdot (\rho \, U_v ) \right], \\
\fl & \qquad \qquad \qquad \quad \qquad \qquad \qquad \quad (\because  \Eref{eqn_rmass2}) \\
\fl &=  \rho  \frac{\partial U_v }{\partial t}  + \kam \left[ \nabla  \left(\nabla \cdot (\rho \, U_v )\right) - \frac{\nabla \rho}{\rho} \nabla \cdot (\rho \, U_v ) \right] - \kam^2 \left[ \nabla \Delta \rho - \frac{\nabla \rho}{\rho}  \Delta \rho \right],}
\end{equation}
\begin{eqnarray}
\label{eqn_C2}
\fl \therefore \rho \, \frac{\partial  U }{\partial t} &&=  \rho  \frac{\partial U_v }{\partial t}  + \kam \left[ \nabla  \left(\nabla \cdot (\rho \, U_v )\right) - \frac{\nabla \rho}{\rho} \nabla \cdot (\rho \, U_v ) \right] - \kam^2 \left[ \nabla \Delta \rho - \frac{\nabla \rho}{\rho}  \Delta \rho \right].
\end{eqnarray}

Similarly, the second term on L.H.S of \eref{eqn_C1} can be transformed into
\begin{equation}
\eqalign{ \fl \rho \, (U \cdot \nabla) \,U = \rho \, \left( U_v \,-\, \kam \,\nabla \ln \rho \right) \cdot \nabla \left( U_v \,-\, \kam\, \nabla \ln \rho \right),\\
\fl = \rho\, (U_v \cdot \nabla) U_v - \rho  \kam \,(U_v \cdot \nabla) \nabla \ln \rho - \rho \kam\,  (\nabla \ln \rho \cdot \nabla) U_v + \rho\, \kam^2 (\nabla \ln \rho \cdot \nabla) \nabla \ln \rho,\\
\fl = \rho\, (U_v \cdot \nabla) U_v \,-\, \kam \left[ (\nabla \rho \cdot \nabla) U_v \,+\, \rho\, (U_v \cdot \nabla) \frac{\nabla \rho}{\rho} \right] \,+\, \kam^2\, (\nabla \rho \cdot \nabla) \frac{\nabla \rho}{\rho},}
\end{equation}

\begin{equation}
\label{eqn_C3}
\fl \eqalign{ \therefore \rho  (U \cdot \nabla) U &= \rho \,(U_v \cdot \nabla) U_v - \kam \left[ (\nabla \rho \cdot \nabla) U_v + \rho \,(U_v \cdot \nabla) \frac{\nabla \rho}{\rho} \right] + \kam^2 (\nabla \rho \cdot \nabla) \frac{\nabla \rho}{\rho}.}
\end{equation}

The momentum balance equation in terms of the volume velocity can be transformed to the following form:
\begin{eqnarray*}
\fl \rho  \frac{\partial U_v }{\partial t}  &&+ \kam \left[ \nabla \left(\rho \left(\nabla \cdot U_v \right) \right) - \nabla \rho \left(\nabla \cdot U_v \right) +\nabla  (U_v \cdot \nabla \rho) - \frac{\nabla \rho}{\rho} (U_v \cdot \nabla \rho) \right]  \nonumber \\
\fl &&  - \kam^2 \left[ \nabla \Delta \rho - \frac{\nabla \rho}{\rho}  \Delta \rho \right] + \rho (U_v \cdot \nabla) U_v - \kam \left[ (\nabla \rho \cdot \nabla) U_v + \rho (U_v \cdot \nabla) \frac{\nabla \rho}{\rho}  \right] \nonumber \\
\fl && + \kam^2 (\nabla \rho \cdot \nabla) \frac{\nabla \rho}{\rho} + \nabla \cdot \left[ p \bI + \bPi_v \right] = 0,
\end{eqnarray*}
or
\begin{eqnarray*}
\rho  \frac{\partial U_v }{\partial t} &&+ \rho (U_v \cdot \nabla) U_v + \kam \rho\, \nabla \left( \nabla \cdot U_v \right) - \kam \left[ \nabla \rho \cdot \nabla U_v - \widetilde{\nabla U_v} \cdot \nabla \rho  \right] \nonumber \\
&&  - \kam^2 \left[ \nabla \Delta \rho - \frac{\nabla \rho}{\rho}  \Delta \rho - (\nabla \rho \cdot \nabla) \frac{\nabla \rho}{\rho}\right] + \nabla \cdot \left[ p \bI + \bPi_v \right] = 0,
\end{eqnarray*}
or
\begin{eqnarray*}
\rho  \frac{\partial U_v }{\partial t} &&+ \rho (U_v \cdot \nabla) U_v - \kam \left[ \nabla \rho \cdot \nabla U_v - \widetilde{\nabla U_v} \cdot \nabla \rho  - \rho\, \nabla \left( \nabla \cdot U_v \right) \right] \nonumber \\
&&  - \kam^2 \left[ \nabla \Delta \rho - \frac{\nabla \rho}{\rho}  \Delta \rho - (\nabla \rho \cdot \nabla) \frac{\nabla \rho}{\rho}\right] + \nabla \cdot \left[ p \bI + \bPi_v \right] = 0.
\end{eqnarray*}

Using \eref{eqn_A4}, we observe that the divergence of $\frac{\nabla \rho \otimes \nabla \rho }{\rho}$ can be written as
\begin{equation}
\label{eqn_C4}
\fl \eqalign{ \nabla \cdot \left[\nabla \rho \otimes \frac{\nabla \rho}{\rho} \right] &= \frac{\nabla \rho}{\rho}  (\nabla \cdot \nabla \rho) + (\nabla \rho \cdot \nabla) \frac{\nabla \rho}{\rho} = \frac{\nabla \rho}{\rho} \Delta \rho + (\nabla \rho \cdot \nabla) \frac{\nabla \rho}{\rho}.}
\end{equation}

With identity given in \eref{eqn_C4}, the final form of the re-casted momentum balance equation is
\begin{eqnarray}
\label{eqn_C5}
\rho  \frac{\partial U_v }{\partial t} && + \rho (U_v \cdot \nabla) U_v + \nabla \cdot \left[ p \bI + \bPi_v \,+\,\kam^2\,\nabla \cdot \left( \nabla \rho \otimes \frac{\nabla \rho}{\rho} \right)\right] \,\nonumber \\
&& -\, \kam \,\left[ \left(\nabla U_v \,-\, \widetilde{\nabla U_v}\right) \cdot \nabla \rho  \,-\, \rho\, \nabla \left( \nabla \cdot U_v \right) \right] \,-\, \kam^2\, \nabla \Delta \rho = 0.
\end{eqnarray}
\section{Re-casting the energy balance equation}
\label{Appendix_D}
Consider the energy balance equation given by \eref{eqn_energy}
\begin{eqnarray}
\label{eqn_D1}
\fl \frac{\partial}{\partial t} \left[\frac{1}{2} \rho U^2 + \rho \, \be_{in}\right] \, +\,\nabla \cdot \left[\frac{1}{2} \rho U^2 U + \rho \, \be_{in} U \right]\,&&+\,\nabla \cdot \left[(p \pmb{I} + \bPi^{(NS)}) \cdot U \right] \nonumber \\
\fl \qquad \qquad \qquad \qquad \qquad \qquad \qquad \qquad  && +\, \nabla \cdot \bq^{(NS)} = 0.
\end{eqnarray}
Our aim is to recast the above energy balance equation which is initially derived in terms of the fluid mass velocity $U$ into an equation in terms of the fluid volume velocity $U_v$.\\

By using the expression for $U^2$ given in \eref{eqn_A8}, the first term in the energy balance equation \eref{eqn_D1} becomes
\begin{eqnarray}
\label{eqn_D2}
\fl \frac{\partial}{\partial t} \left[\frac{1}{2} \rho U^2 \,+\, \rho\, \be_{in}\right] &=& \frac{\partial}{\partial t} \left[\frac{1}{2} \rho U_v^2 + \rho \, \be_{in} \,- \kam\, \rho\, U_v \cdot \nabla \ln \rho \,+\, \frac{1}{2} \kam^2 \nabla \rho \cdot \nabla \ln \rho \right], \nonumber\\
\fl &=& \frac{\partial}{\partial t} \left[\frac{1}{2} \rho U_v^2 + \rho \, \be_{in}\right] \,- \, \kam  \frac{\partial}{\partial t} \left[\rho \, U_v \cdot \nabla \ln \rho \right] \,\nonumber \\
\fl && \quad +\, \frac{1}{2}\, \kam^2\, \frac{\partial}{\partial t} \left[ \nabla \rho \cdot \nabla \ln \rho \right].
\end{eqnarray}

Consider the term $\frac{\partial}{\partial t} \left[\rho \, U_v \cdot \nabla \ln \rho \right]$
\begin{eqnarray}
\label{eqn_D3}
\fl \frac{\partial}{\partial t} \left[\rho \, U_v \cdot \nabla \ln \rho  \right] &=& \frac{\partial (\rho U_{v})}{\partial t} \cdot \nabla \ln \rho  \,+ \, \rho \, U_v \cdot \frac{\partial }{\partial t} \nabla \ln \rho, \nonumber \\
\fl &=& \frac{\partial (\rho U_{v})}{\partial t} \cdot \nabla \ln \rho   \,+ \,  \rho \, U_v \cdot \nabla \left[  \frac{\partial }{\partial t} \ln \rho \right], \nonumber \\
\fl &=& \frac{\partial (\rho U_{v})}{\partial t} \cdot \nabla \ln \rho   \,+ \, U_v \cdot \left[ \nabla \left( \frac{\partial \rho}{\partial t} \right) \,-\,\nabla \ln \rho \, \frac{\partial \rho}{\partial t} \right].
\end{eqnarray}
Using \eref{eqn_B8}, the expression for $\frac{\partial}{\partial t} \left[\rho \, U_v \cdot \nabla \ln \rho \right]$ becomes
\begin{eqnarray}
\label{eqn_D4}
\fl && \frac{\partial}{\partial t} \left[\rho \, U_v \cdot \nabla \ln \rho \right]  = \Bigg \{ - \, \nabla \cdot \left[ \rho \,U_v \otimes U_v\,+\, p \, \bI \,+\, \ \bPi^{(RNS)} \right]\,- \, \kam \, \nabla \left[ \nabla \cdot (\rho \, U_v ) \right]\,\nonumber \\
\fl && \qquad \qquad \qquad \qquad +\, \kam^2  \nabla \Delta \rho \,\Bigg\} \cdot \nabla \ln \rho  \,+\, U_v \cdot \Big[ \nabla \ln \rho\,\nabla \cdot (\rho\, U_v) \,-\, \nabla \left(\nabla \cdot (\rho\, U_v) \right) \,\nonumber \\
\fl && \qquad \qquad \qquad \qquad -\, \kam \, \Delta \rho\, \nabla \ln \rho \, +\, \kam\, \nabla \Delta \rho  \Big].
\end{eqnarray}

Expression for $\frac{\partial}{\partial t} \left[ \nabla \rho \cdot \nabla \ln \rho \right]$:
\begin{eqnarray*}
\fl \frac{\partial}{\partial t} \left[ \nabla \rho \cdot \nabla \ln \rho \right] &=& \left( \frac{\partial}{\partial t}  \nabla \rho \right) \cdot \nabla \ln \rho \,+\, \nabla \rho \cdot \left( \frac{\partial}{\partial t} \nabla \ln \rho \right)  \nonumber \\
\fl &=& \nabla \left(\frac{\partial \rho}{\partial t}\right) \cdot \nabla \ln \rho  + \nabla \rho \cdot \nabla\left[ \frac{1}{\rho} \frac{\partial \rho}{\partial t} \right], \nonumber \\
\fl &=& \nabla \left(\frac{\partial \rho}{\partial t}\right) \cdot \nabla \ln \rho + \nabla \rho \cdot \left[ \frac{-1}{\rho^2} \nabla \rho \frac{\partial \rho}{\partial t} + \frac{1}{\rho} \nabla \frac{\partial \rho}{\partial t} \right], \nonumber \\
\fl &=& \nabla \left(\frac{\partial \rho}{\partial t}\right) \cdot \nabla \ln \rho - \frac{1}{\rho^2} \nabla\rho \cdot \nabla \rho \frac{\partial \rho}{\partial t} + \frac{1}{\rho}  \nabla \rho \cdot \nabla \frac{\partial \rho}{\partial t},  \nonumber \\
\fl &=& 2\, \nabla \left(\frac{\partial \rho}{\partial t}\right) \cdot \nabla \ln \rho  - \frac{|\nabla\rho|^2}{\rho^2} \frac{\partial \rho}{\partial t}, \nonumber \\
\fl &=& 2 \,\nabla \left[- \nabla \cdot \left(\rho\, U_v \right)\, + \,\kam \Delta \rho \right] \cdot \nabla \ln \rho - \frac{|\nabla\rho|^2}{\rho^2} \left[- \nabla \cdot \left(\rho\, U_v \right)\, + \,\kam \Delta \rho \right].
\end{eqnarray*}

\begin{eqnarray}
\label{eqn_D5}
\fl \therefore \, \frac{\partial}{\partial t} \left[ \nabla \rho \cdot \nabla \ln \rho \right]&=& - 2 \, \nabla \left[\nabla \cdot (\rho\, U_v) \right]  \cdot \nabla \ln \rho \, +\, 2 \,\kam \,\nabla \Delta \rho \cdot \nabla \ln \rho \,\nonumber \\
&& + \frac{|\nabla\rho|^2}{\rho^2} \nabla \cdot \left(\rho\, U_v \right)\,-\,\frac{\kam\, |\nabla\rho|^2}{\rho^2}  \Delta \rho.
\end{eqnarray}

Finally, using \eref{eqn_D4} and \eref{eqn_D5} in \eref{eqn_D2} we have
\begin{eqnarray}
\fl  \frac{\partial}{\partial t} \left[\frac{1}{2} \rho U^2 + \rho \, \be_{in}\right] = \frac{\partial}{\partial t} \left[\frac{1}{2} \rho U_v^2 + \rho \, \be_{in}\right] \,+\, \kam\, \nabla \cdot \left[ \rho \,U_v \otimes U_v\,+\, p \, \bI \,+\, \ \bPi^{(RNS)} \right] \cdot \nabla \ln \rho \, \nonumber \\
\fl  + \kam^2 \, \nabla \left[ \nabla \cdot (\rho \, U_v ) \right] \cdot \nabla \ln \rho \,-\, \kam \, U_v \cdot \Big[ \nabla \ln \rho\,\nabla \cdot (\rho\, U_v) \,-\, \nabla \left(\nabla \cdot (\rho\, U_v) \right) \Big]\, \nonumber \\
\fl  -\, \kam^3\,  \nabla \Delta \rho \,\cdot \nabla \ln \rho\, + \, \kam^2 \left( U_v \cdot \Delta \rho\, \nabla \ln \rho \right)\, -\, \kam^2 \left( U_v \cdot \nabla \Delta \rho \right) \,- \,\kam^2\, \nabla \left[\nabla \cdot (\rho\, U_v) \right]  \cdot \nabla \ln \rho \,\nonumber \\
\fl + \,\kam^3 \,\nabla \Delta \rho \cdot \nabla \ln \rho \,+ \frac{1}{2} \kam^2 \frac{|\nabla\rho|^2}{\rho^2} \nabla \cdot \left(\rho\, U_v \right)\,-\,\frac{1}{2} \kam^3\, \frac{|\nabla\rho|^2}{\rho^2}  \Delta \rho,
\end{eqnarray}
or
\begin{eqnarray}
\label{eqn_D6}
\fl \frac{\partial}{\partial t} \left[\frac{1}{2} \rho U^2 + \rho \, \be_{in}\right] = \frac{\partial}{\partial t} \left[\frac{1}{2} \rho U_v^2 + \rho \, \be_{in}\right] \,+\, \kam\, \nabla \cdot \left[ \rho \,U_v \otimes U_v\,+\, p \, \bI \,+\, \ \bPi^{(RNS)} \right] \cdot \nabla \ln \rho \, \nonumber \\
\fl \qquad \qquad  -\, \kam \, U_v \cdot \Big[ \nabla \ln \rho\,\nabla \cdot (\rho\, U_v) \,-\, \nabla \left(\nabla \cdot (\rho\, U_v) \right) \Big]\, + \, \kam^2 \left( U_v \cdot \Delta \rho\, \nabla \ln \rho \right)\, \nonumber \\
\fl \qquad \qquad -\, \kam^2 \left( U_v \cdot \nabla \Delta \rho \right)\,+ \frac{1}{2} \kam^2 \frac{|\nabla\rho|^2}{\rho^2} \nabla \cdot \left(\rho\, U_v \right)\,-\,\frac{1}{2} \kam^3\, \frac{|\nabla\rho|^2}{\rho^2}  \Delta \rho,
\end{eqnarray}
or
\begin{eqnarray}
\label{eqn_D7}
\fl \frac{\partial}{\partial t} \left[\frac{1}{2} \rho U^2 + \rho \, \be_{in}\right] &=& \frac{\partial}{\partial t} \left[\frac{1}{2} \rho U_v^2 + \rho \, \be_{in}\right] \,-\, \kam^3\, \frac{|\nabla\rho|^2}{2\,\rho^2}  \Delta \rho\, \nonumber \\
\fl &&   +\,\kam^2\, \Bigg\{\left( U_v \cdot \Delta \rho\, \nabla \ln \rho \right)\, -\, \left( U_v \cdot \nabla \Delta \rho \right)\,+\, \frac{1}{2} \frac{|\nabla\rho|^2}{\rho^2} \nabla \cdot \left(\rho\, U_v \right) \Bigg\} \, \nonumber \\
\fl &&  -\, \kam\, \Bigg\{ U_v \cdot \Big[ \nabla \ln \rho\,\nabla \cdot (\rho\, U_v) \,-\, \nabla \left(\nabla \cdot (\rho\, U_v) \right) \Big] \,\nonumber \\
\fl && \qquad \qquad - \, \nabla \cdot \left[ \rho \,U_v \otimes U_v\,+\, p \, \bI \,+\, \ \bPi^{(RNS)} \right] \cdot \nabla \ln \rho  \Bigg\}.
\end{eqnarray}

Now, let us consider the second term $\nabla \cdot \left[\frac{1}{2} \rho U^2 U + \rho e_{in} U \right]$ in the energy equation \eref{eqn_C1}. In terms of the new velocity field $U_v$, the expression $\frac{1}{2} \rho U^2 U + \rho e_{in} U $ is transformed to
\begin{eqnarray}
\label{eqn_D8}
\fl \frac{1}{2}\, \rho \,U^2 \,U\, +\, \rho\, \be_{in}\, U = \frac{1}{2} \rho \,U_v^2 \,U_v \,+ \,\rho\, \be_{in}\, U_v \, - \kam\, \rho \, \be_{in}\, \nabla \ln \rho \,-\, \kam \,(U_v \cdot \nabla \rho)\, U_v \,\nonumber \\
\fl \qquad \quad -\,  \frac{1}{2}\, \kam \,U_v^2 \nabla \rho \, -\,  \frac{\kam^3}{2\, \rho}  |\nabla \rho|^2 \, \nabla \ln \rho  \,+\, \kam^2\, (U_v \cdot \nabla \rho) \, \nabla \ln \rho \,+\, \frac{1}{2} \frac{\kam^2 }{\rho}\, |\nabla \rho|^2 \, U_v.
\end{eqnarray}

Furthermore, we observe that
\begin{eqnarray}
\label{eqn_D9}
\left(p\,\bI\,+\, \bPi^{(NS)} \right) \cdot U &=& \left( p\,\bI\, +\,\bPi_v \right) \cdot \left(U_v\,- \, \kam \,\nabla \ln \rho\right)\nonumber \\
&=& \left( p\,\bI\, +\,\bPi_v\right) \cdot U_v\,- \kam \,\left( p\,\bI\, +\,\bPi_v\right) \cdot \nabla \ln \rho.
\end{eqnarray}
Finally, one can write the re-casted energy balance equation using \eref{eqn_D7}, \eref{eqn_D8} and \eref{eqn_D9}. The energy balance equation in terms of the volume velocity $U_v$ is then,
\begin{eqnarray}
\label{eqn_D10}
\fl && \frac{\partial}{\partial t} \left[\frac{1}{2} \rho U_v^2 + \rho \, \be_{in}\right] \,- \, \kam  \frac{\partial}{\partial t} \left[\rho \, U_v \cdot \nabla \ln \rho \right] \,+\, \frac{1}{2}\, \kam^2\, \frac{\partial}{\partial t} \left[ \nabla \rho \cdot \nabla \ln \rho \right]\,\nonumber \\
\fl && +\,\nabla \cdot \Bigg[ \frac{1}{2} \rho \,U_v^2 \,U_v \,+ \,\rho\, \be_{in}\, U_v \, - \kam\, \rho \, \be_{in}\, \nabla \ln \rho \,-\, \kam \,(U_v \cdot \nabla \rho)\, U_v \,-\,  \frac{1}{2}\, \kam \,U_v^2 \nabla \rho \,\nonumber \\
\fl && \qquad \qquad -\,  \frac{\kam^3}{2\, \rho}  |\nabla \rho|^2 \, \nabla \ln \rho  \,+\, \kam^2\, (U_v \cdot \nabla \rho) \, \nabla \ln \rho \,+\, \frac{1}{2} \frac{\kam^2 }{\rho}\, |\nabla \rho|^2 \, U_v \Bigg] \nonumber \\
\fl && +\,\nabla \cdot \Bigg[ \left( p\,\bI\, +\,\bPi_v \right) \cdot \left(U_v\,- \, \kam \,\nabla \ln \rho\right) \Bigg]\, +\,\nabla \cdot \left[ \bq^{(NS)}\right] = 0,
\end{eqnarray}
or
\begin{eqnarray}
\label{eqn_D11}
\fl && \frac{\partial}{\partial t} \left[\frac{1}{2} \rho U_v^2 + \rho \, \be_{in}\right] \,+\,\nabla \cdot \left[ \frac{1}{2} \rho \,U_v^2 \,U_v \,+ \,\rho\, \be_{in}\, U_v \right]\,+\,\nabla \cdot \Bigg[ \left( p\,\bI + \bPi_v \right) \cdot U_v \Bigg]\nonumber \\
\fl && +\,\nabla \cdot \Bigg[- \kam \,\bPi_v \cdot \nabla \ln \rho \Bigg] + \nabla \cdot \left[ \bq^{(NS)} - \kam\, \rho \, \be_{in}\, \nabla \ln \rho - \kam\,p\,\bI \cdot \nabla \ln \rho \right] \nonumber \\
\fl && + \nabla \cdot \Bigg[-\, \kam \,(U_v \cdot \nabla \rho)\, U_v -  \frac{1}{2}\, \kam \,U_v^2 \nabla \rho \,-\,  \frac{\kam^3}{2\, \rho}  |\nabla \rho|^2 \, \nabla \ln \rho  \,+\, \kam^2\, (U_v \cdot \nabla \rho) \, \nabla \ln \rho \,\nonumber \\
\fl && \qquad \quad +\, \frac{1}{2} \frac{\kam^2 }{\rho}\, |\nabla \rho|^2 \, U_v \Bigg]\,- \, \kam  \frac{\partial}{\partial t} \left[\rho \, U_v \cdot \nabla \ln \rho \right] \,+\, \frac{1}{2}\, \kam^2\, \frac{\partial}{\partial t} \left[ \nabla \rho \cdot \nabla \ln \rho \right]= 0,
\end{eqnarray}
or
\begin{eqnarray}
\label{eqn_D12}
\fl && \frac{\partial}{\partial t} \left[\frac{1}{2} \rho U_v^2 + \rho \, \be_{in}\right] \,+\,\nabla \cdot \left[ \frac{1}{2} \rho \,U_v^2 \,U_v \,+ \,\rho\, \be_{in}\, U_v \right]\,+\,\nabla \cdot \Bigg[ \left( p\,\bI\, +\,\bPi_v \right) \cdot U_v \Bigg]\nonumber \\
\fl && -\,\nabla \cdot \Bigg[ \kam \,\bPi_v \cdot \nabla \ln \rho \Bigg] \,+\,\nabla \cdot \left[ \bq^{(NS)}\,-\, \kam\, \rho \, \be_{in}\, \nabla \ln \rho \,-\, \kam\,p\,\bI \cdot \nabla \ln \rho \right] \nonumber \\
\fl && +\,\nabla \cdot \Bigg[  \,-\, \kam \,(U_v \cdot \nabla \rho)\, U_v \,-\,  \frac{1}{2}\, \kam \,U_v^2 \nabla \rho \,-\,  \frac{\kam^3}{2\, \rho}  |\nabla \rho|^2 \, \nabla \ln \rho  \,+\, \kam^2\, (U_v \cdot \nabla \rho) \, \nabla \ln \rho \, \nonumber \\
\fl && \qquad \quad +\, \frac{1}{2} \frac{\kam^2 }{\rho}\, |\nabla \rho|^2 \, U_v \Bigg]\, -\, \kam\, \Bigg\{ U_v \cdot \Big[ \nabla \ln \rho\,\nabla \cdot (\rho\, U_v) \,-\, \nabla \left(\nabla \cdot (\rho\, U_v) \right) \Big] \, \nonumber \\
\fl && \qquad \qquad \qquad \qquad \qquad \qquad  - \nabla \cdot \left[ \rho \,U_v \otimes U_v\,+\, p \, \bI \,+\, \ \bPi^{(RNS)}_v \right] \cdot \nabla \ln \rho  \Bigg\}\,\nonumber \\
\fl && + \kam^2\, \Bigg[\left( U_v \cdot \Delta \rho\, \nabla \ln \rho \right) - \left( U_v \cdot \nabla \Delta \rho \right) + \frac{1}{2} \frac{|\nabla\rho|^2}{\rho^2} \nabla \cdot \left(\rho\, U_v \right) \Bigg] - \kam^3\, \frac{|\nabla\rho|^2}{2\,\rho^2}  \Delta \rho\,= 0.
\end{eqnarray}

Rearranging the terms in the above equation, we obtain the final form of the re-casted energy balance equation as
\begin{eqnarray}
\label{eqn_D13}
\fl \frac{\partial}{\partial t} \left[\frac{1}{2} \rho U_v^2 + \rho \, \be_{in}\right] \,+\,\nabla \cdot \left[ \frac{1}{2} \rho \,U_v^2 \,U_v \,+ \,\rho\, \be_{in}\, U_v \right]\,+\,\nabla \cdot \Bigg[ \left( p\,\bI\, +\,\bPi_v \right) \cdot U_v\,\Bigg]\nonumber \\
\fl -\,\nabla \cdot \Bigg[ \kam \,\bPi_v \cdot \nabla \ln \rho \Bigg] +\,\nabla \cdot \left[ \bq^{(NS)}\,-\, \kam\,\left( \rho \, \be_{in}\, \nabla \ln \rho \,+\,p\,\bI \cdot \nabla \ln \rho \right) \right] \nonumber \\
\fl  +\,\nabla \cdot \Bigg\{\kam \,\left[ -\,(U_v \cdot \nabla \rho)\, U_v \,-\,  \frac{1}{2}\, U_v^2 \,\nabla \rho \right] \,+\,\kam^2\, \left[ (U_v \cdot \nabla \rho) \, \nabla \ln \rho \,+\, \frac{1}{2\,\rho}\, |\nabla \rho|^2 \, U_v \right] \,\nonumber \\
\fl  \qquad \qquad \qquad \qquad +\,\kam^3\,\left[- \frac{1}{2\, \rho}  |\nabla \rho|^2 \, \nabla \ln \rho \right]\Bigg\} \nonumber \\
\fl + \kam\, \Bigg\{ \nabla \cdot \left[ \rho \,U_v \otimes U_v + p \, \bI + \bPi^{(RNS)}_v \right] \cdot \nabla \ln \rho \nonumber \\
\fl \qquad \qquad - U_v \cdot \Big[\nabla \ln \rho\,\nabla \cdot (\rho\, U_v) - \nabla \left(\nabla \cdot (\rho\, U_v) \right) \Big]  \Bigg\}\,\nonumber \\
\fl  +\,\kam^2\, \Bigg[\left( U_v \cdot \Delta \rho\, \nabla \ln \rho \right)\, -\, \left( U_v \cdot \nabla \Delta \rho \right) \,+\, \frac{1}{2} \frac{|\nabla\rho|^2}{\rho^2} \nabla \cdot \left(\rho\, U_v \right) \Bigg]\,\nonumber \\
\fl +\, \kam^3\, \left[-\,\frac{1}{2\,\rho^2} |\nabla\rho|^2\, \Delta \rho \right]\,= 0. 
\end{eqnarray}
Let us assume that
\begin{eqnarray}
\fl \mcn_{v_1} &=& -\,(U_v \cdot \nabla \rho)\, U_v \,-\,  \frac{1}{2}\, U_v^2 \,\nabla \rho, \\
\fl \mcn_{v_2} &=& (U_v \cdot \nabla \rho) \, \nabla \ln \rho \,+\, \frac{1}{2\,\rho}\, |\nabla \rho|^2 \, U_v, \\
\fl \mcn_{v_3} &=& - \frac{1}{2\, \rho}  |\nabla \rho|^2 \, \nabla \ln \rho, \\
\fl \mcn_{v_4} &=& \nabla \cdot \left[ \rho \,U_v \otimes U_v\,+\, p \, \bI \,+\, \ \bPi^{(RNS)}_v \right] \cdot \nabla \ln \rho \,\nonumber \\
\fl && \qquad \qquad -\,U_v \cdot \Big[ \nabla \ln \rho\,\nabla \cdot (\rho\, U_v) \, -\, \nabla \left(\nabla \cdot (\rho\, U_v) \right) \Big],\\
\fl \mcn_{v_5} &=& \left( U_v \cdot \Delta \rho\, \nabla \ln \rho \right)\, -\, \left( U_v \cdot \nabla \Delta \rho \right) \,+\, \frac{1}{2} \frac{|\nabla\rho|^2}{\rho^2} \nabla \cdot \left(\rho\, U_v \right), \\
\fl \mcn_{v_6} &=& -\,\frac{1}{2\,\rho^2} |\nabla\rho|^2\, \Delta \rho,
\end{eqnarray}
then the final form of the re-casted energy balance equation is given by

\begin{eqnarray}
\label{eqn_D14}
\fl  \frac{\partial}{\partial t} \left[\frac{1}{2} \rho U_v^2 + \rho \, \be_{in}\right] \,&&+\,\nabla \cdot \left[ \frac{1}{2} \rho \,U_v^2 \,U_v \,+ \,\rho\, \be_{in}\, U_v \right]\,\nonumber \\
\fl && +\,\nabla \cdot \Bigg[ \left( p\,\bI\, +\,\bPi_v \right) \cdot U_v\,- \, \kam \,\bPi_v \cdot \nabla \ln \rho \Bigg] \, \nonumber \\
\fl && +\,\nabla \cdot \left[ \bq^{(NS)}\,-\, \kam\,\Big( \rho \, \be_{in}\, \nabla \ln \rho \,+\,p\,\bI \cdot \nabla \ln \rho \Big) \right] \, \nonumber \\
\fl && +\,\nabla \cdot \Bigg[\kam \,\mcn_{v_1} \,+\,\kam^2\, \mcn_{v_2} \,+\,\kam^3\,\mcn_{v_3} \Bigg] \nonumber \\
\fl && +\, \kam\, \mcn_{v_4}\,+\,\kam^2\, \mcn_{v_5}\,+\, \kam^3\, \mcn_{v_6}\,= 0.
\end{eqnarray}
\section*{References}
\bibliography{ReddyDadzieetal_Paperarxiv_09Sep2019}

\end{document}